\documentclass[floatfix,showpacs,superscriptaddress,prl]{revtex4}
\usepackage{amsbsy,amssymb,amsmath,bm}
\usepackage{float}
\usepackage{epsf}
\usepackage{graphicx,color}
\usepackage[caption = false]{subfig}
\usepackage{amsfonts}
\usepackage{amssymb}
\usepackage{amsmath}
\usepackage[T1]{fontenc}
\usepackage{xspace}
\usepackage{epsfig}
\usepackage{ctable}
\newcommand{\rrangle}{>\kern-1.2ex~>\xspace}

\definecolor{red}{rgb}{1,0,0}           
\definecolor{green}{rgb}{0,1,0}
\definecolor{blue}{rgb}{0,0,1}
\definecolor{darkblue}{rgb}{0,0,0.5}
\definecolor{lightblue}{rgb}{.5,.5,1}
\definecolor{lightgray}{gray}{.87}          
\definecolor{Dark}{gray}{.20}
\definecolor{pink}{rgb}{.95,0.82,0.92}  
\definecolor{yellow}{rgb}{1,1,0}
\definecolor{lightyellow}{rgb}{1,1,.5}
\definecolor{purple}{rgb}{0.7,0,0.85}
\definecolor{darkgreen}{rgb}{0,0.5,0}
\definecolor{orange}{rgb}{0.8,0.2,0.2}

\def \be {\bea}
\def \ee {\eea}
\def \bea {\begin{eqnarray}}
\def \eea {\end{eqnarray}}
\def \bse {\begin{subequations}}
\def \ese {\end{subequations}}
\def \bde {\begin{description}}
\def \ede {\end{description}}
\def \nn {\nonumber}

\def \la {\langle}
\def \ra {\rangle}

\def \rra {\rangle \hskip-0.2cm \rangle}

\def \pd {\partial}

\def \a {\alpha}
\def \b {\beta}
\def \dag {\dagger}
\def \g {\gamma}

\def \n {\nu}

\def \om {\omega}

\def \t {\tau}

\def \e {\mbox{e}}

\def \tA {\tilde{A}}
\def \tB {\tilde{B}}

\def \HH {\mathcal{H}}

\def \UU  {\mathcal{U}}

\begin{document}

\title{Connection between the winding number and the Chern number }

\author{Han-Ting Chen}
\affiliation{\textit{Physics Department, National Taiwan University, Taipei 10617, Taiwan}}
\author{Chia-Hsun Chang}
\affiliation{\textit{Physics Department, National Taiwan Normal University, Taipei 11677, Taiwan}}
\author{Hsien-chung Kao\footnote{e-mail address: hckao@ntnu.edu.tw}}
\affiliation{\textit{Physics Department, National Taiwan Normal University, Taipei 11677, Taiwan}}

\date{\today }

\begin{abstract}
Bulk-edge correspondence is one of the most distinct properties of topological insulators. In particular, the 1D winding number $\n$ has a one-to-one correspondence to the number of edge states in a chain of topological insulators with boundaries. By properly choosing the unit cells, we carry out numerical calculation to show explicitly in the extended SSH model that the winding numbers corresponding to the left and right unit cells may be used to predict the numbers of edge states on the two boundaries in a finite chain.  Moreover, by drawing analogy between the SSH model and QWZ model, we show that the extended SSH model may be generalized to the extended QWZ model. By integrating the ``magnetic field'' over the momentum strip $0\le p_2 \le \pi, 0\le p_1 2\pi$ in the Brillouin zone, we show a identity relating the 2D Chern number and the difference between the 1D winding numbers at $p_2=0 $ and $p_2 =\pi$. 
\end{abstract}

\maketitle

\subsection{I. Introduction}

Since the discovery of topological materials,  they have drawn a lot of attention in the community of condensed matter physicists \cite{Review}.  Bulk-edge correspondence is one of the most distinct properties of topological insulators. In particular, the one-dimension (1D) winding number $\n$ has a one-to-one correspondence to the number of edge states in a chain of topological insulators with boundaries.  When there is only a single connected boundary between the topological material and the environment, it is quite straight forward to make sense of the correspondence.  On the other hand, if there are two or more disconnected boundaries, then it is sometimes not so easy to correctly interpret the results if we look into the details.  For simplicity, let's consider a finite chain of the SSH model, which is one of the simplest topological materials\cite{SSH, Rice-Mele}. The number of edge states depends on whether the total number of sites in the system is even or odd. If the chain is in the topological phase and the number of sites is even, then there will be two edge states, which seems plausible since there are two boundaries after all. However, if the number of sites is odd, then there is always one edge state, which appears on either the left or right boundary depending on whether the inter-cell hopping amplitude is larger or smaller than the intra-cell hopping amplitude. Hence, it is not so transparent why this happens.

It is also known in the literature that we may extend the SSH model to the Rice-Mele model by adding an on-site energy term \cite{Rice-Mele}. Making use of the system, we may relate the SSH model to a Chern insulator by consider a charge-pumping process in the system. In fact, it has been shown that the 2D Chern number is equal to the number of particles pumped from the left to the right boundaries in a cycle \cite{Thouless, polarization}. Although this does provide some physical insight into our understanding of the 2D Chern number, it would be even better if we can establish an identity that directly relates the two topological invariants. Moreover, such an identity may also help shed more light upon why the Zak phase is generally not quantized in the Rice-Mele model \cite{Zak, Zak-phase_quantization}.

In this paper, we try to address the above two problems. After an in-depth analysis, we put forward some possible resolutions. The rest of the paper is organized in the following way. In Sec. II, we first use the SSH model to carry out a detailed analysis of the bulk-edge correspondence. In a finite chain of SSH model, there are generally two boundaries, the left and right ones. To make the bulk-edge correspondence work sensibly, it is shown that we must choose the unit cells in such a way that they are consistent with the left and right boundaries, respectively. The winding numbers $\n$ corresponding to the two unit cells may then be used to predict the numbers of edge states on the left and right boundaries. Then, we show that the bulk-edge correspondence would also work in the extended SSH model in which there are also next to nearest neighbor hopping amplitudes so that the highest winding number becomes 2. We make use of the results to understand the edge states in carbon nanotubes with various edges. It is demonstrated that the existence of edge states depend sensitively on the boundary conditions. In Sec. III, we use the SSH model and the corresponding Chern insulator to establish a relation between the 2D Chern number and the difference of the 1D winding numbers at $p_2 =0$ and $p_2 =\pi$. We then demonstrate that this relation also hold for the extended SSH models. We expect similar identity would also exist in higher dimensions. Finally, we make conclusion and discuss possible extensions in Sec. IV.

\subsection{II. The winding number and the bulk-edge correspondence}

Let's begin with the well-known SSH model, whose Hamiltonian is given by
\bea
H_{\rm SSH}=\sum_{j=-\infty}^{\infty}\left\{ \left(t_0  A_j^\dag  +t_1 A_{j+1}^\dag\right) B_j \right\}  + {\rm h.c.}.
\eea
Here, $j$ denotes the unit cell, and $t_0, t_1$ are the intra-cell and inter-cell hopping amplitudes, respectively.  Without loss of generality, we will assume $t_0, t_1$ to be both positive through out the paper for convenience. The Bloch Hamiltonian takes the following form
\be\label{SSH}
\HH_{\rm SSH}=\left(
\begin{matrix}
0 & h^*(p) \cr
h(p) & 0 \cr
\end{matrix}
\right),
\ee
with  $h(p)=t_0+t_1 \e^{ip}$. In terms of the Pauli matrices $\tau_i$'s, we have:
\bea
\HH_{\rm SSH}= \left(t_0 + t_1\cos p\right)\tau_1 + \left(t_1\sin p \right)\tau_2.
\eea
It is obvious that the chiral operator $\Pi = \tau_3$ anti-commutes with $\HH_{\rm SSH}$, which means that eigenstates of $\HH_{\rm SSH}$ with non-zero energy always appear in pairs with eigenvalues $(E, -E)$ and the corresponding eigenstates are related by $\left|-E \right\ra = \tau_3 \left|E \right\ra$. In contrast, zero energy eigenstates can always be chosen to be chiral eigenstates and the left-handed and right-handed states are decoupled from each other.

Whether the system is in the topological phase or not may be determined by the 1D winding number $\n$ derived from $h(p)$, which traces out a closed contour in the complex plane as $p$ ranges over the Brillouin zone. It is well-known that the analytical expression for $\n$ is given by
\bea
\n = \frac{-i}{2\pi} \int_{0}^{2\pi} dp \frac{h'(p)}{h(p)}= \frac{1}{2\pi} \int_{0}^{2\pi} dp \left\{ \frac{e^{ip}}{e^{ip} - \mathfrak{s}}\right\} , \label{1d-winding-number}
\eea
where $ \mathfrak{s}=-t_0/t_1$. When $t_1>t_0$ we have $|\mathfrak{s}|<1$. In this case, the contour will encircle the origin once so that $\n=1$, and the system is in the topological phase. In contrast, when $t_1<t_0$ we have  $|\mathfrak{s}|>1$. Now, the origin will not be enclosed by the contour so that $\n=0$, and the system is in the trivial phase.  According to the so-called "bulk-edge correspondence", the most salient signature for a system to be in the topological phase is the appearance of zero energy edge states on the boundaries of the system.  However, it is also well-known that there is a freedom in choosing the unit-cell in the SSH model.  Rather than grouping $A_j$ and $B_j$ to be the $j$-th unit-cell, we may rename $B_j$ and $A_{j+1}$ to be $\tA_j$ and $\tB_j$ for example and group them into a unit-cell instead. It is obvious that when we do this, the role of $t_0$ and $t_1$ will be interchanged and a system that is classified to be topological would become trivial under the new choice of unit-cell and vice versa. Naturally this leads to the question that if there is an ambiguity in determining whether a system is in the topological phase or not, then how we would be able to make sense of the bulk-edge correspondence? In order to resolve this difficulty, let's first consider a right semi-infinite SSH chain:
\bea
H^{\rm R}_{\rm SSH}=\sum_{j=1}^{\infty}\left\{ \left(t_0  A_j^\dag  +t_1 A_{j+1}^\dag\right) B_j  \right\} + {\rm h.c.},
\eea
so that $A_1$ is the site by the left edge of the system. The energy eigenstates would satisfy the following recurrence relation and boundary condition:
\bea \label{R-semi-inf}
&\;& \hskip -3.1cm E A_j - \left(t_0 B_j +t_1 B_{j-1} \right) =0; \cr
&\;& \hskip -3.1cm E B_j - \left(t_0 A_j +t_1 A_{j+1} \right) =0, \cr
&\;& \hskip -3.1cm B_0 =0.
\eea
An edge state would be described by
\bea \label{R-semi-inf edge-state}
&\;& \hskip -3.1cm A_j =\a s^j, B_j=\b s^j.
\eea
By substituting the above expression into Eq. (\ref{R-semi-inf}), we have
\bea
&\;& \hskip -3.1cm E \a - \left(t_0 +t_1 s^{-1} \right)\b =0; \cr
&\;& \hskip -3.1cm E \b - \left(t_0 +t_1 s \right)\a =0; \cr
&\;& \hskip -3.1cm \b=0.
\eea
To be consistent with the boundary condition, we see from the above equation that a non-trivial solution exists only if
\bea
E=0, {\rm and}\; s= -t_0/t_1.
\eea
For the wave function to be normalizable, we must have $|s|<1$.  This implies $t_1>t_0$ and
\bea
A_j = A_1\left(-t_0/t_1 \right)^{j-1},\;  B_j=0, {\rm with}\; j\ge 1.
\eea

In contrast, if we consider a right semi-infinite SSH chain starting at $B_0$, i.e. adding one more $B$ site to left of the previous chain, then the Schrodinger equation would remain the same but the boundary condition would become
\bea \label{R-semi-inf2}
&\;& \hskip -3.1cm A_0 =0.
\eea
It is obvious that now the corresponding non-trivial solution is given by
\bea
B_{j-1}= B_0\left(-t_1/t_0 \right)^{j},\;  A_j =0, {\rm with}\; j\ge 1.
\eea
Thus, it exists only if $t_0>t_1$.  This result would also be consistent with the bulk-edge correspondence if we now choose to group $B_j$ and $A_{j+1}$ into a unit-cell instead. In this case, the roles of the intra-cell and inter-cell hopping amplitudes are interchanged. Analogous analysis may be carried out for a left semi-infinite SSH chain and similar conclusion may be drawn. In a nutshell, we must choose the unit-cell properly according to the boundary of the system for the bulk-edge correspondence to work sensibly.  A zero energy edge state would show up whenever the inter-cell hopping amplitude is larger than that of the intra-cell.

\begin{figure}[h]
\centering
  \begin{minipage}[b]{0.5\textwidth}
   \includegraphics[width=0.7\textwidth]{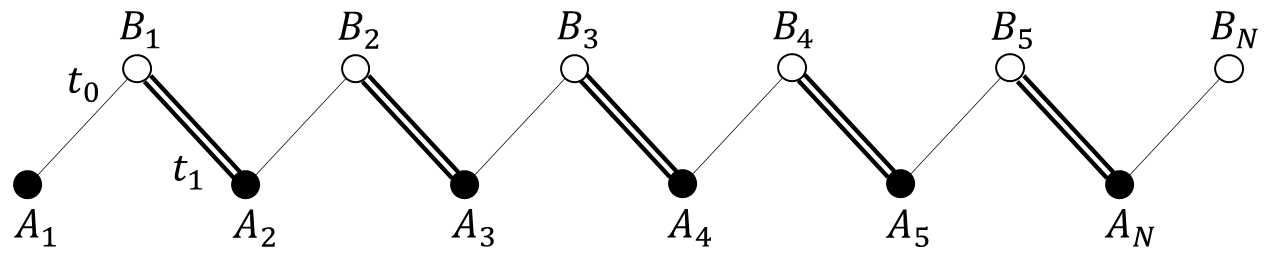}
  \end{minipage}
  \hspace{1cm} 
  \begin{minipage}[b]{0.5\textwidth}
    \includegraphics[width=0.7\textwidth]{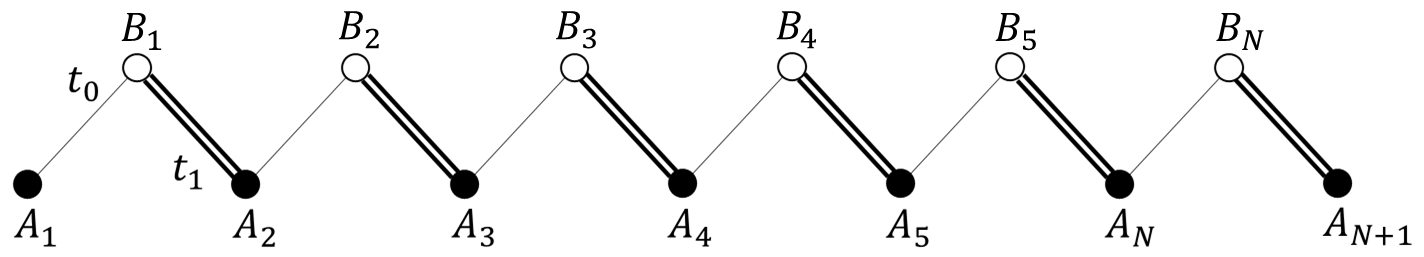}
  \end{minipage}
\caption{The SSH system with $2N$ and $2N+1$ particles.}
    \label{FIG finite ssh chain}
\end{figure}

We may make further check on this conclusion by considering a finite chain of the SSH model (see Fig.~\ref{FIG finite ssh chain}). In such a case, there are now two boundaries, and we must choose the unit-cell properly for the left and right boundaries separately.  Let's first consider the simpler case that there are even number of sites:
\bea
&\;& \hskip-1.0cm H^{\rm even}_{\rm SSH}=\sum_{j=1}^{N-1}\left\{  \left(t_0  A_j^\dag  +t_1 A_{j+1}^\dag\right) B_j  \right\} +t_0 A_N^\dag B_N + {\rm h.c.}. \nn \\
&\;& \hskip-0.5cm
\eea

\begin{figure}[h]
\centering
\subfloat[]{\includegraphics[width=0.30\textwidth]{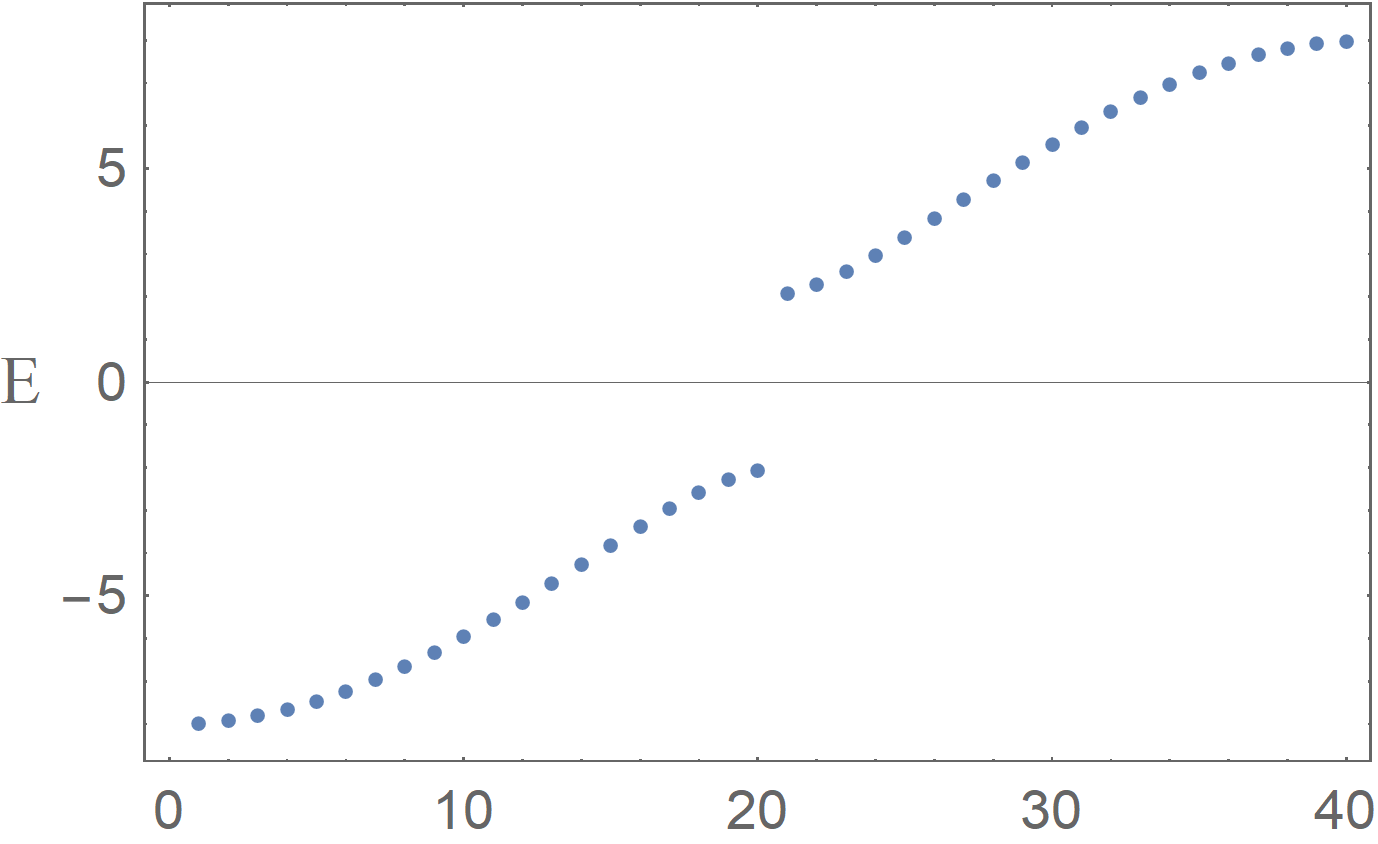}}\hskip 0.5cm
\subfloat[]{\includegraphics[width=0.30\textwidth]{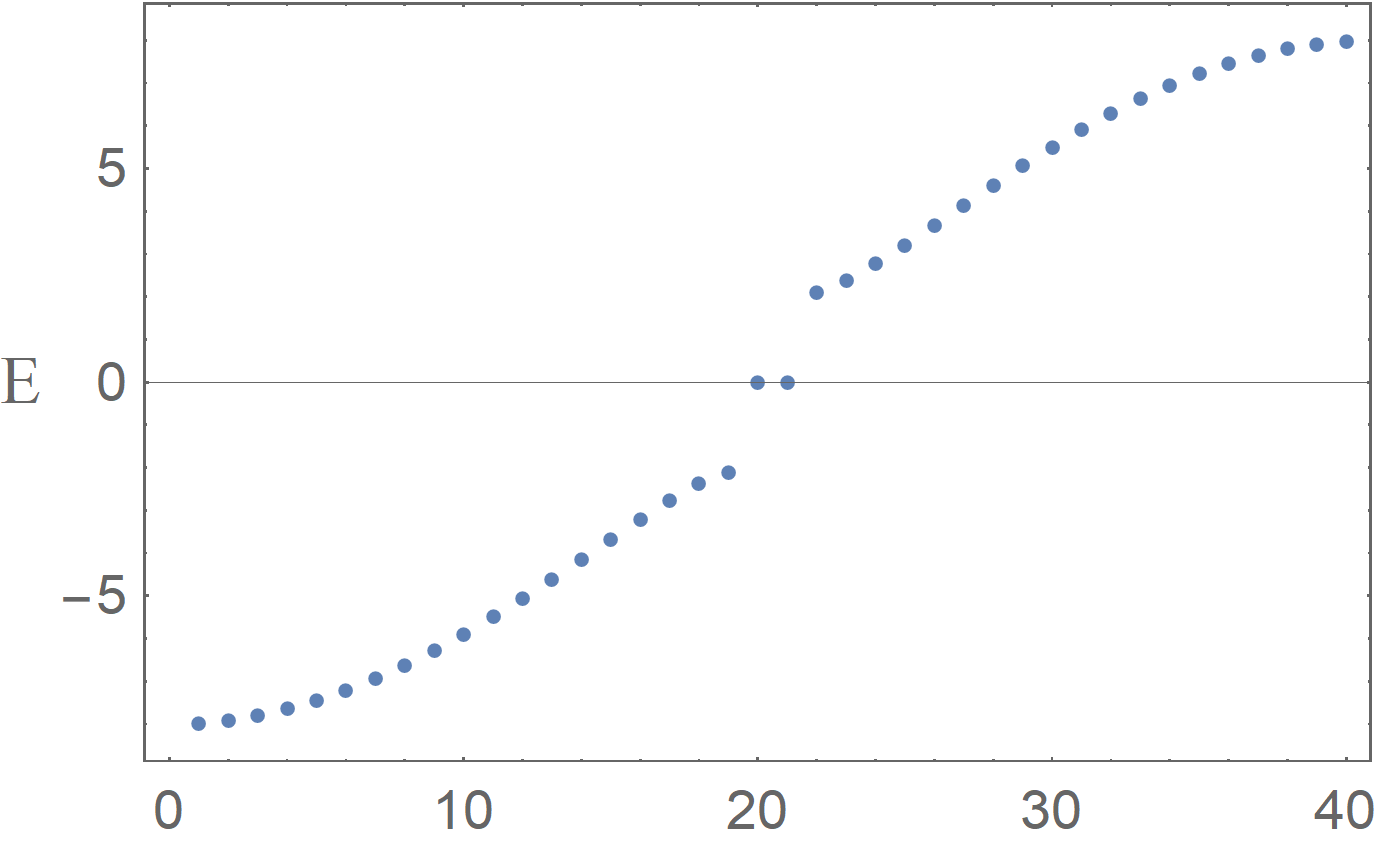}}\\
\subfloat[]{\includegraphics[width=0.30\textwidth]{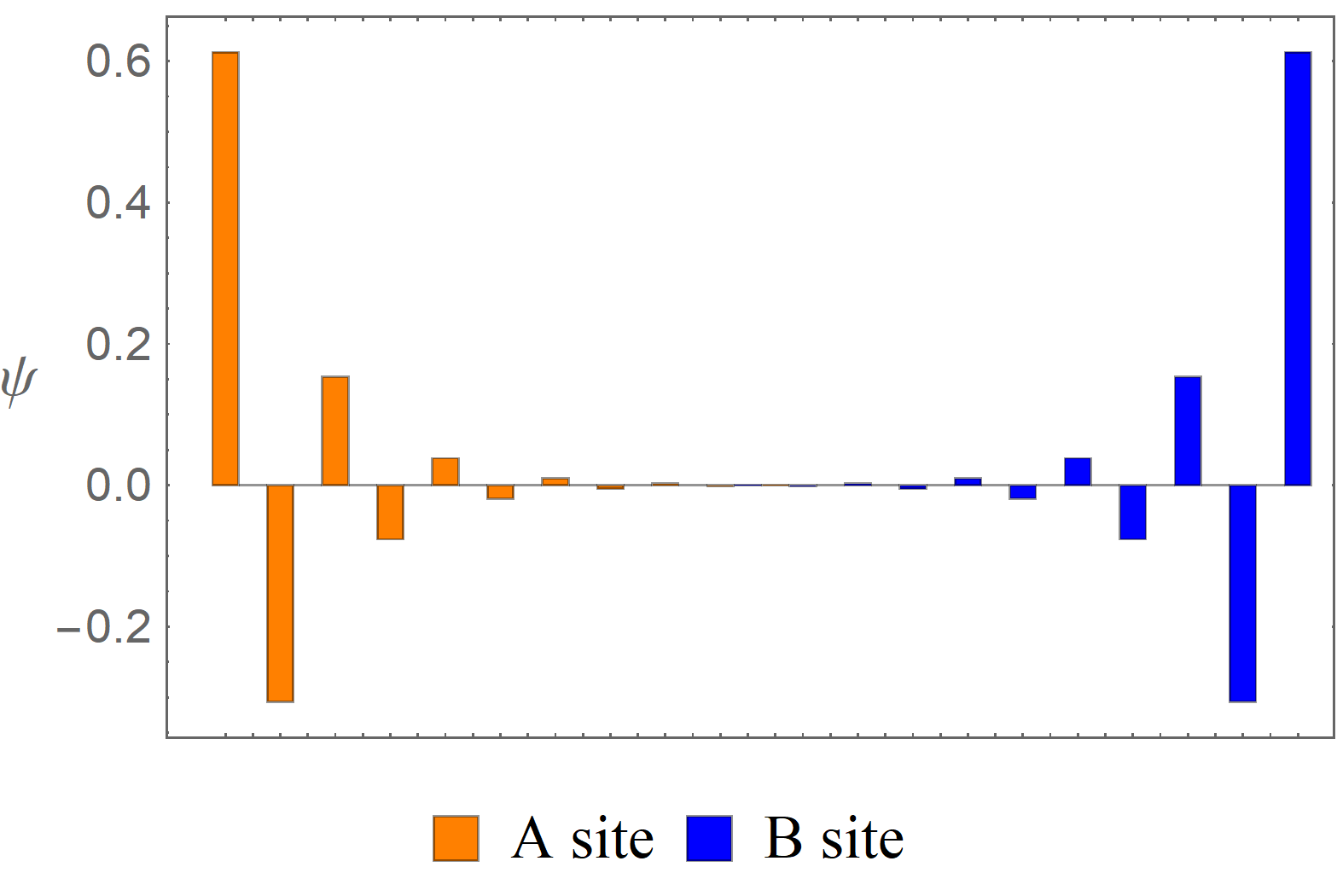}}\hskip 0.5cm
\subfloat[]{\includegraphics[width=0.30\textwidth]{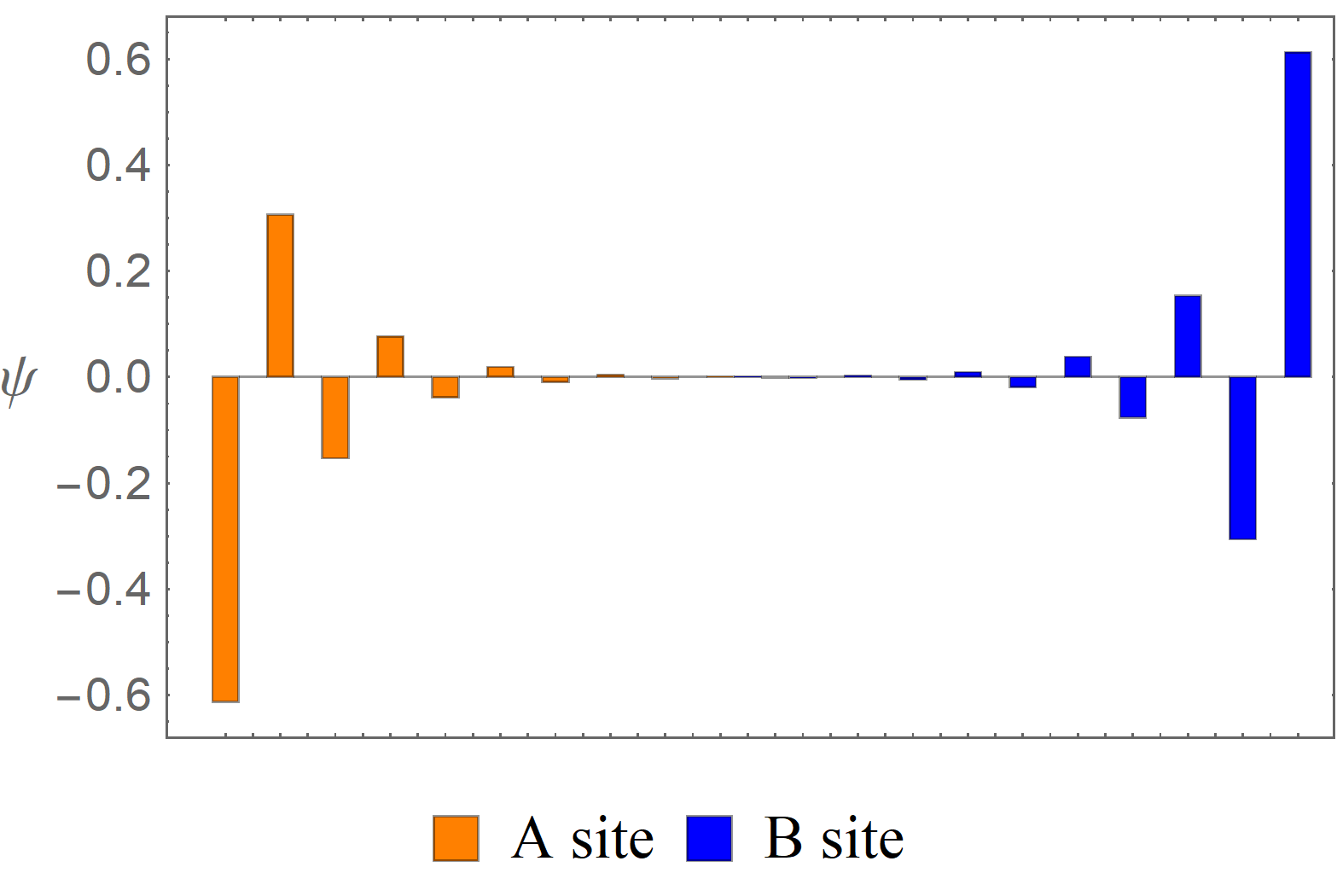}}\\
\subfloat[]{\includegraphics[width=0.32\textwidth]{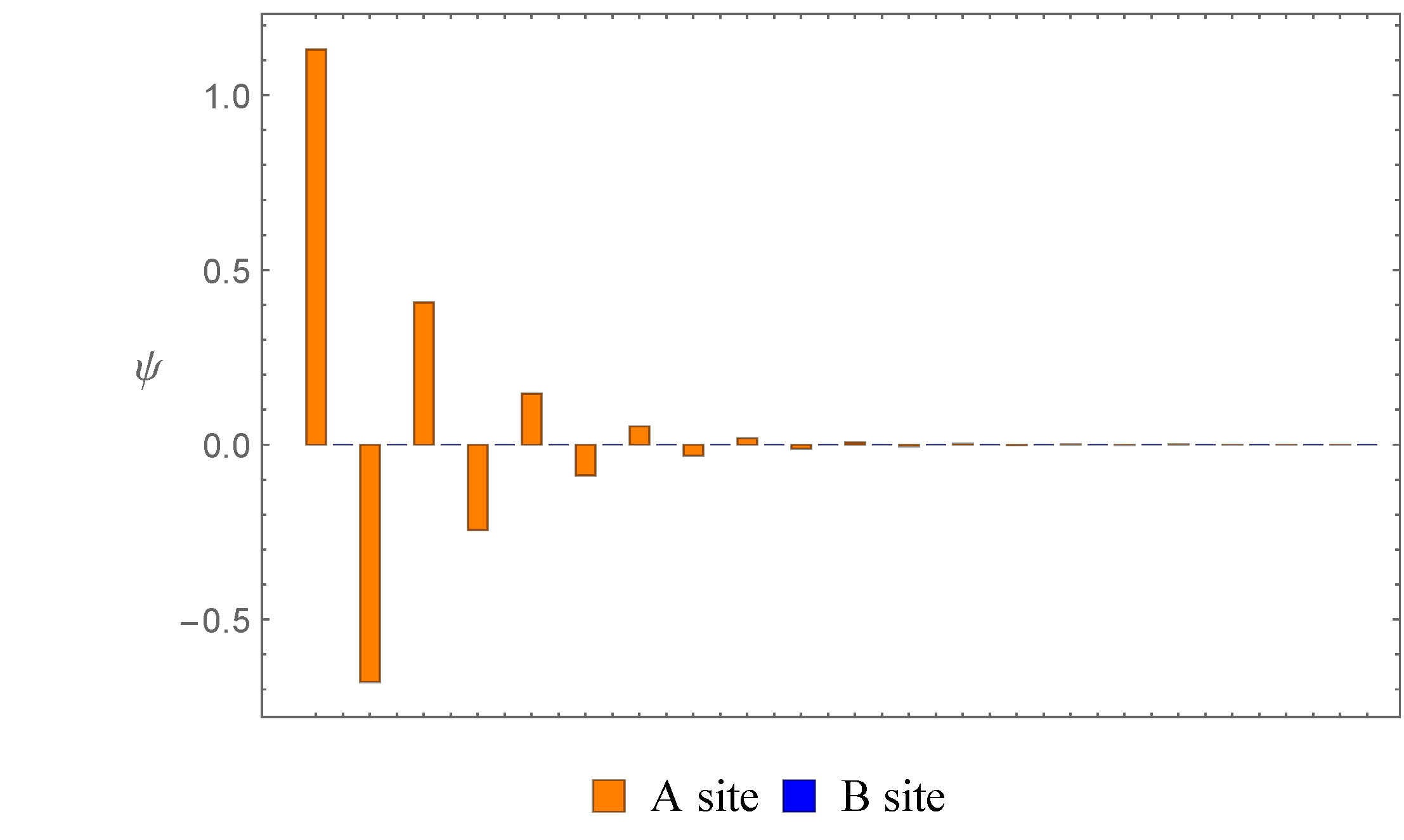}}\hskip 0.5cm
\subfloat[]{\includegraphics[width=0.32\textwidth]{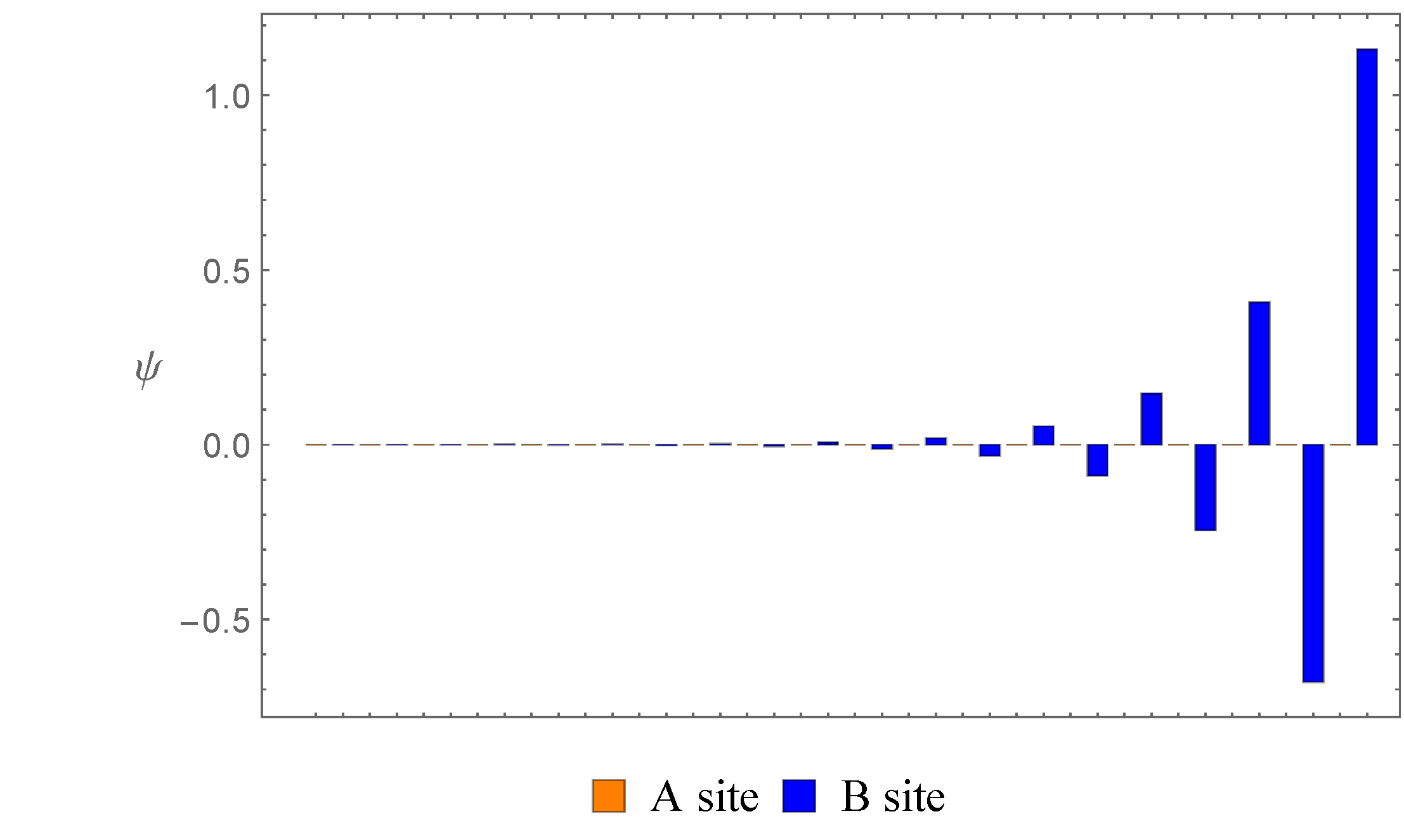}}
\caption{(a) Energy spectrum in the trivial phase with $(t_0,t_1,N)=(5,3,20)$. (b) Energy spectrum in the topological phase with $(t_0,t_1,N)=(3,5,20)$. (c),(d) The wave functions of the almost zero energy states in the topological phase. (e),(f) By taking the sum and difference of the even and odd wave functions of the almost zero energy states in (c) and (d), we arrive at the left and right edge states, respectively. }\label{fig ssh even eigenvector}
\end{figure}

Using the standard technique to solve the recurrence relation, one finds that
\bea
&\;& \hskip -2.1cm E=\pm\sqrt{t_0^2+t_1^2+t_0 t_1\left(s+s^{-1}\right) }; \cr
&\;& \hskip -2.1cm t_0\left(\frac{s^{N+1}-s^{-N-1}}{s-s^{-1}} \right) +t_1 \left(\frac{s^{N}-s^{-N}}{s-s^{-1}} \right)=0. \label{eqs}
\eea
In this case, the solutions can only be found numerically. Because of the configurations of the two boundaries, it is obvious that we should group $A_j$ and $B_j$ into a unit-cell for both of them. Based on the experience we obtained for a semi-infinite chain, we expect that edge states would exist only if $t_1>t_0$. Since the left and right edges are now separated by a finite distance, there would be mixing between the two edge states due to quantum tunneling. Consequently, the energy of these edge states would be approximately zero and the corresponding solutions to Eq.~(\ref{eqs}) are $s\approx -t_0/t_1, -t_1/t_0$. This indeed may be explicitly verified by numerical calculation, which is shown in Fig.~\ref{fig ssh even eigenvector}.  It can be seen that the wave functions in  Figs.~\ref{fig ssh even eigenvector}c and  \ref{fig ssh even eigenvector}d are even and odd functions of the position, respectively.  By taking the sum and difference of these two states, one may see explicitly that the resulting states become the left and right edge states, which are non-vanishing only on sites $A$ and $B$, respectively, just as we expected. As we know edge states decay exponentially away from the boundaries, we can find numerically the best fit to the base number of the exponential function: $-0.5947$ and $1.682$. For sure, they are well consistent with the values $-t_0/t_1$ and $ -t_1/t_0$. Therefore, it is in perfect agreement with the bulk-edge correspondence.

\begin{figure}[h]
\centering
  \begin{minipage}[b]{0.5\textwidth}
   \includegraphics[width=0.7\textwidth]{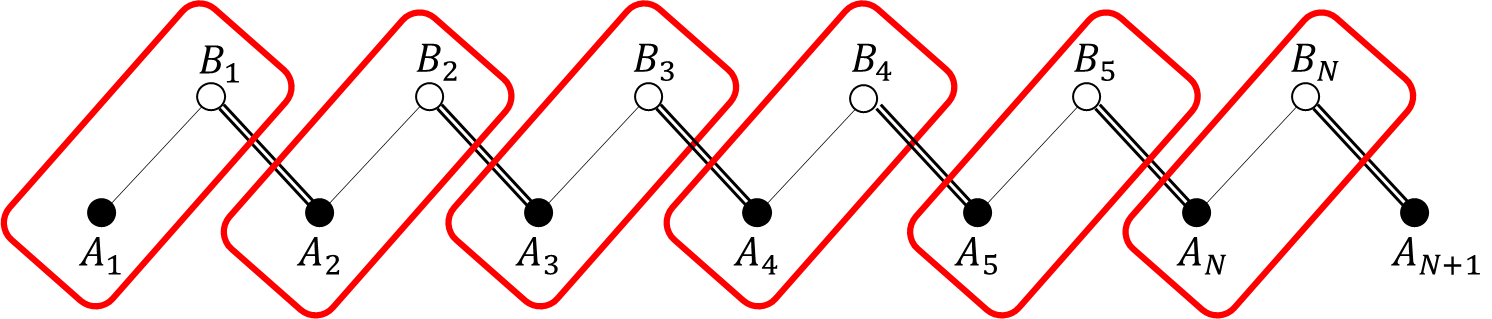}
  \end{minipage}
  \hspace{1cm} 
  \begin{minipage}[b]{0.5\textwidth}
    \includegraphics[width=0.7\textwidth]{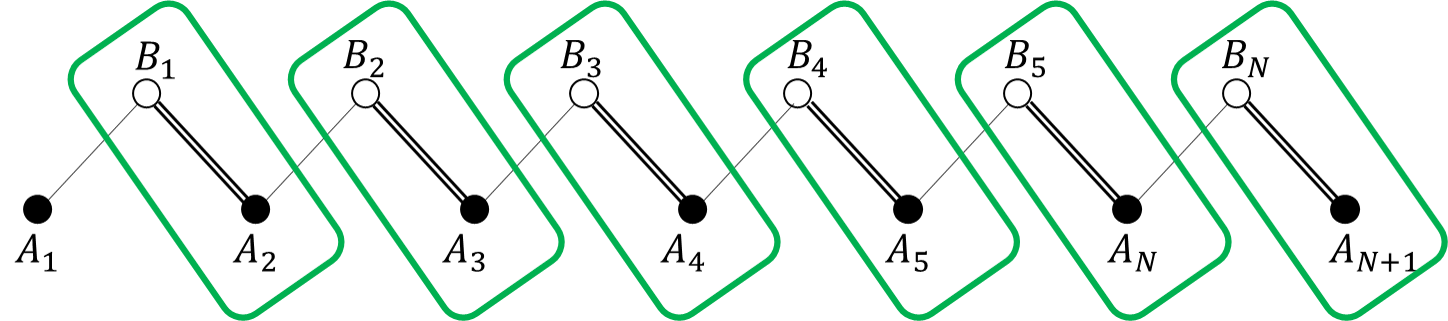}
  \end{minipage}
\caption{Two possible ways to choose the unit cell, none of which can cover the whole system. Top: The unit cell that is consistent with the left edge of the system. Bottom: The unit cell that is consistent with the right edge.}\label{fig ssh odd chain left/right cell}
\end{figure}

On the other hand, when there are odd number of sites in the system, we have
\bea
H^{\rm odd}_{\rm SSH}=\sum_{j=1}^{N}\left\{ \left(t_0  A_j^\dag  +t_1 A_{j+1}^\dag\right) B_j  \right\} + {\rm h.c.}.
\eea
In this case, the recursion relation may be solved analytically:
\bea
&\;& \hskip -6.1cm s=e^{i\pi k/N},
\eea
with $k=1,\ldots, N,$ and
\bea
&\;& \hskip -2.1cm E_k =\pm\sqrt{t_0^2+t_1^2+2t_0 t_1\cos\left[\pi k/(N+1)\right] }.
\eea
In this case, the form of the chiral symmetry operator is modified to
\be
 \Pi=\left(
\begin{matrix}
\tau_3 &0        &\ldots & \ldots  & 0 \cr
0        &\tau_3 & 0       & \ldots  & 0 \cr
0        &\ldots  &\ddots & \ldots  & 0 \cr
0        &\ldots   & 0      &  \tau_3 & 0 \cr
0        &\ldots   &\ldots  & 0         &1  \cr
\end{matrix}
\right).
\ee
In addition to the above $N$ pairs of energy eigenstates, there is also an edge state with exact zero energy dictated by the chiral symmetry of the system.  Here and later on, we may find the corresponding $s=-t_0/t_1$ by using the chiral projection \cite{chiral zero modes}. Note that when the total number of sites is odd, the unit-cell consistent with the left boundary is always different from the one that is consistent with the right boundary (see Fig.~\ref{fig ssh odd chain left/right cell}). When $t_1 >t_0$, the unit-cell associated with the left boundary is in the topological phase and the one associated with the right boundary is in the trivial phase.  This again makes perfect sense since it can be seen that there is indeed a left edge state. In contrast, when $t_1<t_0$, the unit-cell associated with the left boundary is in the trivial phase while the one associated with the right boundary is in the topological phase, and thus there is a right edge state. Consequently, the bulk-edge correspondence is again uphold.

The bulk-edge correspondence seen in the SSH model may be generalized to systems with next to nearest neighbor hopping amplitudes, which we call the extended SSH models.  To be more specific, let's  consider the following two types of extended SSH models. The Hamiltonians are given by
\bea\label{ssh ext1 hamiltonian}
&\;& \hskip-0.7cm H_{\mathrm{ext}1}=\underset{j=-\infty}{\overset{\infty}{\sum }}
\left\{\left(t_0 A_{j}^{\dagger} +t_1 A_{j+1}^{\dagger }+ t_2 A_{j+2}^{\dagger } \right)B_{j} + {\rm h.c.}\right\},
\eea
and
\bea\label{ssh ext2 hamiltonian}
&\;& \hskip-0.9cm H_{\mathrm{ext}2}=\underset{j=-\infty}{\overset{\infty}{\sum }} \left\{\left(t_{-1} A_{j-1}^{\dagger}+ t_0 A_{j}^{\dagger } + t_1 A_{j+1}^{\dagger} \right)B_{j} + {\rm h.c.} \right\},
\eea
respectively. The corresponding Bloch Hamiltonians are then given by 
\be\label{Ext-SSH}
\HH_{\rm ext1}=\left(
\begin{matrix}
0 & h_{\rm ext1}^*(p) \cr
h_{\rm ext1}(p) & 0 \cr
\end{matrix}
\right), \mbox{ and }
\HH_{\rm ext2}=\left(
\begin{matrix}
0 & h_{\rm ext2}^*(p) \cr
h_{\rm ext2}(p) & 0 \cr
\end{matrix}
\right).
\ee
Here, 
\bea
&\;& \hskip-3.5cm h_{\mathrm{ext}1}(p)=\left(t_0 + t_1 \e^{ip} + t_2 \e^{2ip} \right),
\eea
and
\bea
&\;& \hskip-3.0cm h_{\mathrm{ext}2}(p)=\e^{-ip}\left(t_{-1} + t_0 \e^{ip} + t_1 \e^{2ip} \right).
\eea
Note that in the type 1 extended SSH model if we rename $B_j$ and $A_{j+1}$ as $\tA_j$ and $\tB_{j}$ so that they are grouped into a unit cell instead, then $H_{\mathrm{ext}1}$ would become
\bea\label{ssh ext1 hamiltonian tilde}
&\;& \hskip-0.7cm \tilde{H}_{\mathrm{ext}1}=\underset{j=-\infty}{\overset{\infty}{\sum }}
\left\{\left(t_0 \tB_{j-1}^{\dagger} +t_1 \tB_{j}^{\dagger }+ t_2 \tB_{j+1}^{\dagger } \right)\tA_{j} + {\rm h.c.}\right\}, \nn \\
&\;&  \hskip0.3cm =\underset{j=-\infty}{\overset{\infty}{\sum }}
\left\{\left(t_0 \tA_{j+1}^{\dagger} +t_1 \tA_{j}^{\dagger }+ t_2 \tA_{j-1}^{\dagger } \right)\tB_{j} + {\rm h.c.}\right\}.
\eea
It is thus equivalent to $H_{\mathrm{ext}2}$ if we rename properly the hopping amplitudes $(t_0 , t_1, t_2)$ as $(\tilde{t}_1 , \tilde{t}_0, \tilde{t}_{-1})$. Consequently, once we learn how to classify the type 1 extended SSH model, it is straight forward to see the corresponding classification of the type 2 extended SSH model and vice versa.

Without lost of generality, we may make $t_1$ and $t_2$ positive and rewrite
\bea
\label{quadratic equation of s}
&\;& \hskip -2.5cm h_{\mathrm{ext}1}(p)=t_2\left(\e^{ip} - \mathfrak{s}_{1} \right)\left(\e^{ip} -\mathfrak{s}_{2} \right),
\eea
with
\bea
\label{solution of quadratic equation}
\mathfrak{s}_{1}=\frac{-t_1+\sqrt{t_1^2-4t_0t_2}}{2t_2},\; \mathfrak{s}_{2}=\frac{-t_1-\sqrt{t_1^2-4t_0t_2}}{2t_2}.
\eea
The expression in Eq.~(\ref{1d-winding-number}) may be easily generalized to the current case, and we have
\bea
\n = \frac{1}{2\pi} \int_{0}^{2\pi} dp \left\{ \frac{\e^{ip}}{\e^{ip} - \mathfrak{s}_{1}} +  \frac{\e^{ip}}{\e^{ip} - \mathfrak{s}_{2}}  \right\}.
\eea
Whenever $|\mathfrak{s}_{i}|<1$,  the point $\mathfrak{s}_{i}$ will be enclosed by the unit circle, and the corresponding integral will contribute a value of 1 to the winding number $\n$.  On the other hand, it is known that the energy eigenstates of the right semi-infinite chain may be found by solving the following recurrence relation and boundary condition
\bea
&\;& \hskip -2.5cm E A_j - \left(t_0 B_j + t_1 B_{j-1} + t_2 B_{j-2}\right) =0; \cr
&\;& \hskip -2.5cm E B_j - \left(t_0 A_j + t_1 A_{j+1} + t_2 A_{j+2}\right) =0, \cr
&\;& \hskip -2.5cm B_0 = B_{-1} =0. \label{next_to_nearest_hopping}
\eea
In particular, it has been shown in Ref.~\cite{chiral zero modes} that the chiral zero modes arise from the solutions to the following characteristic equation
\bea
&\;& \hskip -2.5cm t_0 + t_1 s + t_2 s^2= t_2 \left(s - \mathfrak{s}_{1} \right)\left(s -\mathfrak{s}_{2} \right)=0,
\eea
which satisfy the condition
\bea
&\;& \hskip -7.5cm |\mathfrak{s}_{i}|<1,
\eea
so that the corresponding wave function is normalizable.

From the above analysis, it is again transparent to see the bulk-edge correspondence. To obtain explicitly the relation between the winding number $\nu$ and the parameters of the system $t_0, t_1,$ and $t_2$, we factor $h_{\mathrm{ext}1}(p)$ in the following way:
\bea
h_{\mathrm{ext}1}(p)= \e^{ip} \left(t_2 \e^{ip} +t_1 + t_0 \e^{-ip}\right).
\eea
It is clear that the factor $\e^{ip}$ would always contribute a value of 1 to the winding number.  Meanwhile, since
\bea
&\;& \hskip -2.5cm t_2 \e^{ip} +t_1 + t_0 \e^{-ip} = t_1 +(t_2+t_0)\cos p + i(t_2-t_0)\sin p,
\eea
it traces out an ellipse with center $(t_1, 0)$ on the complex plane.  As a result, the second factor would have vanishing contribution to the winding number if
\bea
|t_2+t_0|<t_1.
\eea
In contrast, if
\bea
|t_2+t_0|>t_1,
\eea
it would contribute 1 and -1 to the winding number for $t_2-|t_0|>0$ and  $t_2-|t_0|<0$, respectively. Thus, we may classify the system according to the winding number and there are three categories:
\bde
\item{i.)} $|\mathfrak{s}_{1}|<1,|\mathfrak{s}_{2}|<1$, ($|t_2+t_0|>t_1$, and  $t_2-|t_0|>0$):
The system is in the topological phase with winding number $\n=2$.
In this case,  there should be two edge states on the corresponding boundary and they are described by
\bea
&\;& \hskip -2.0cm A_{j} = \a_1 s_1^{j} + \a_2 s_2^{j},\; B_j =0.
\eea

\item{ii.)} $|\mathfrak{s}_{1}|<1,|\mathfrak{s}_{2}|>1$,  ($|t_2+t_0|<t_1$):
The system is in the topological phase with $\n=1$.
In this case, there should be one edge state on the corresponding boundary and it is given by
\bea
&\;& \hskip -3.0cm A_{j} = \a_1 s_1^{j},\; B_j =0.
\eea

\item{iii.)} $|\mathfrak{s}_{1}|>1,\mathfrak{s}_{2}|>1$, ($|t_2+t_0|>t_1$, and  $t_2-|t_0|<0$):
The system is in the trivial phase with $\n=0$, and there would be no edge state on the boundary.
\ede

\begin{figure}[hbt!]
\centering
\subfloat[]{\includegraphics[width=0.60\textwidth]{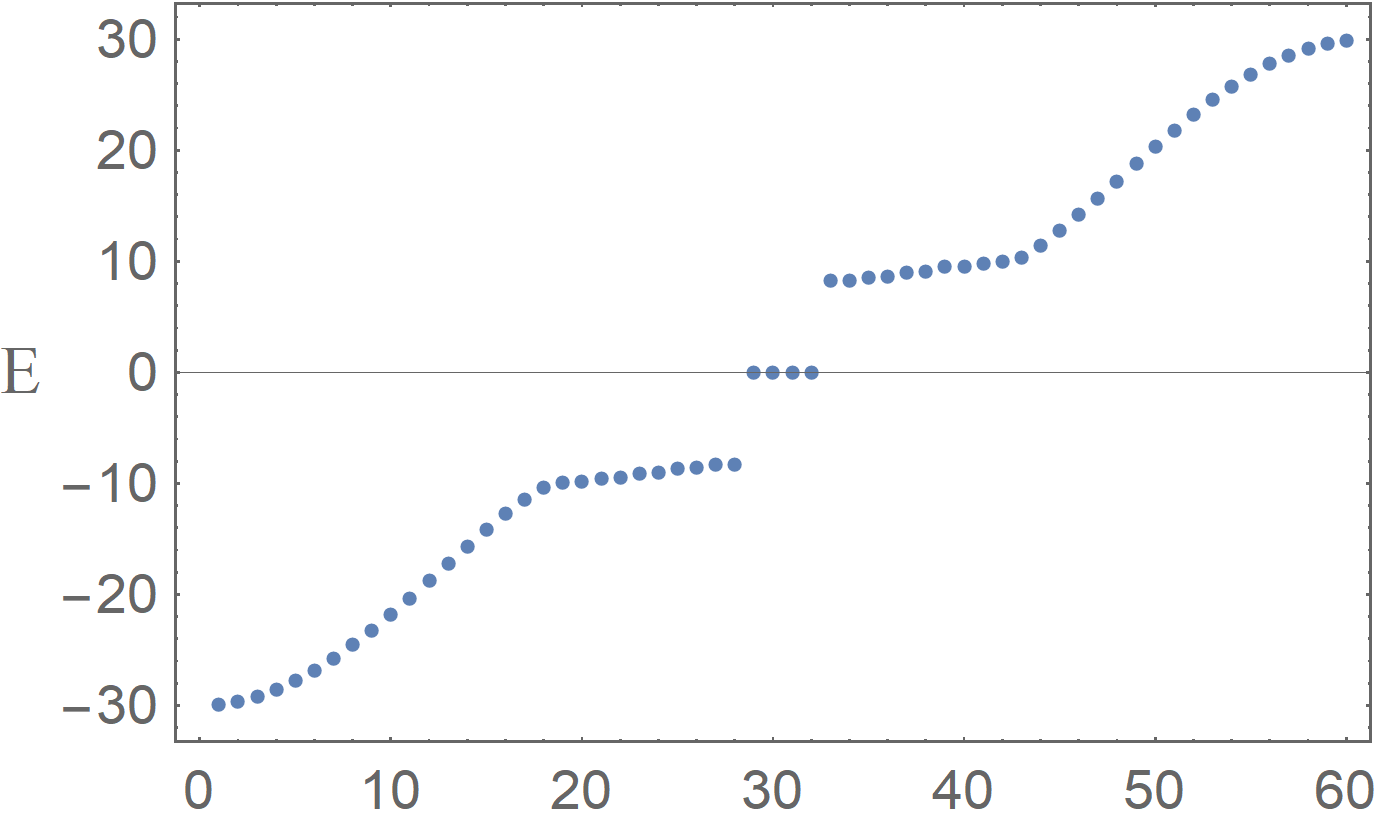}}\\
\subfloat[]{\includegraphics[width=0.30\textwidth]{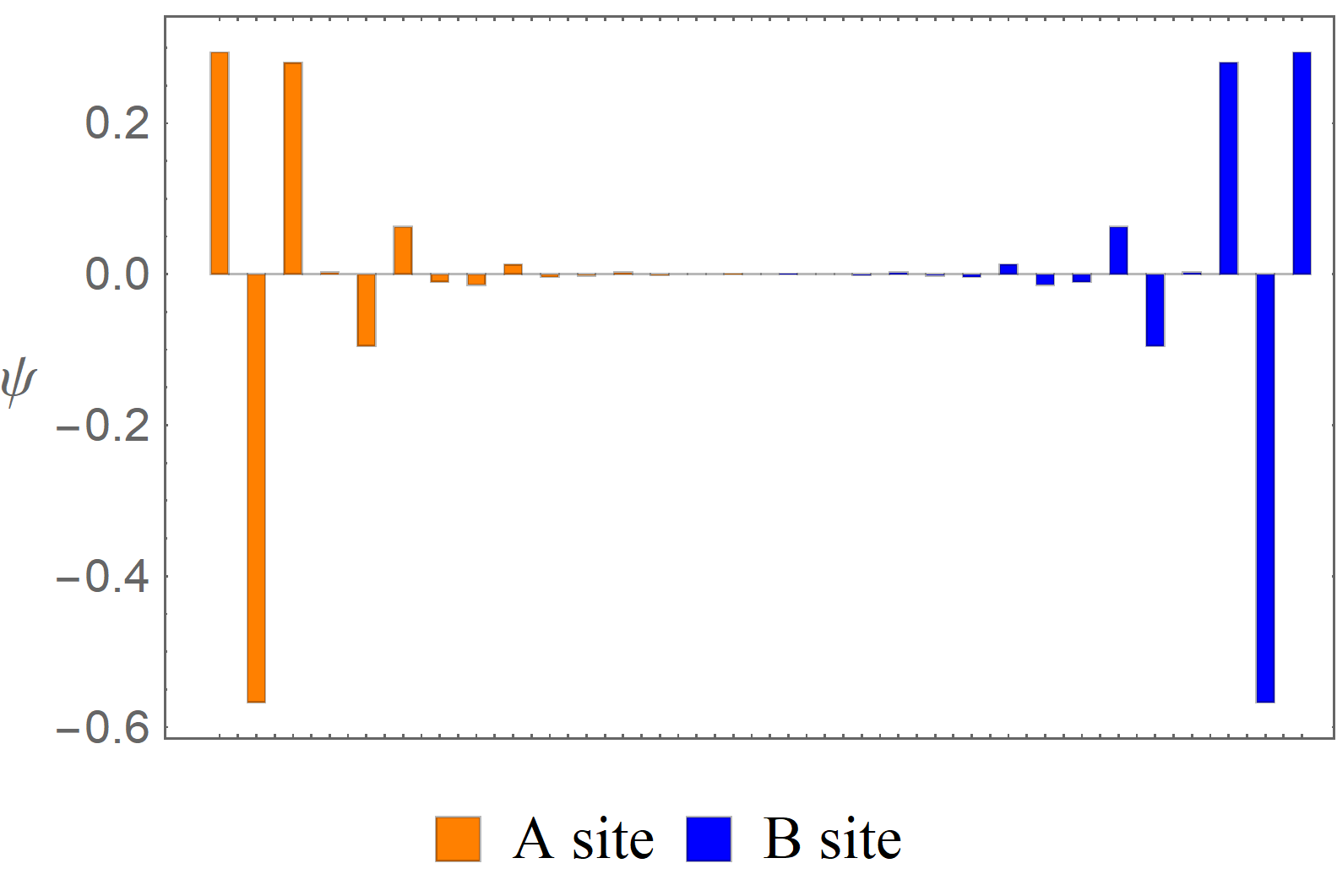}}\hskip 0.5cm
\subfloat[]{\includegraphics[width=0.30\textwidth]{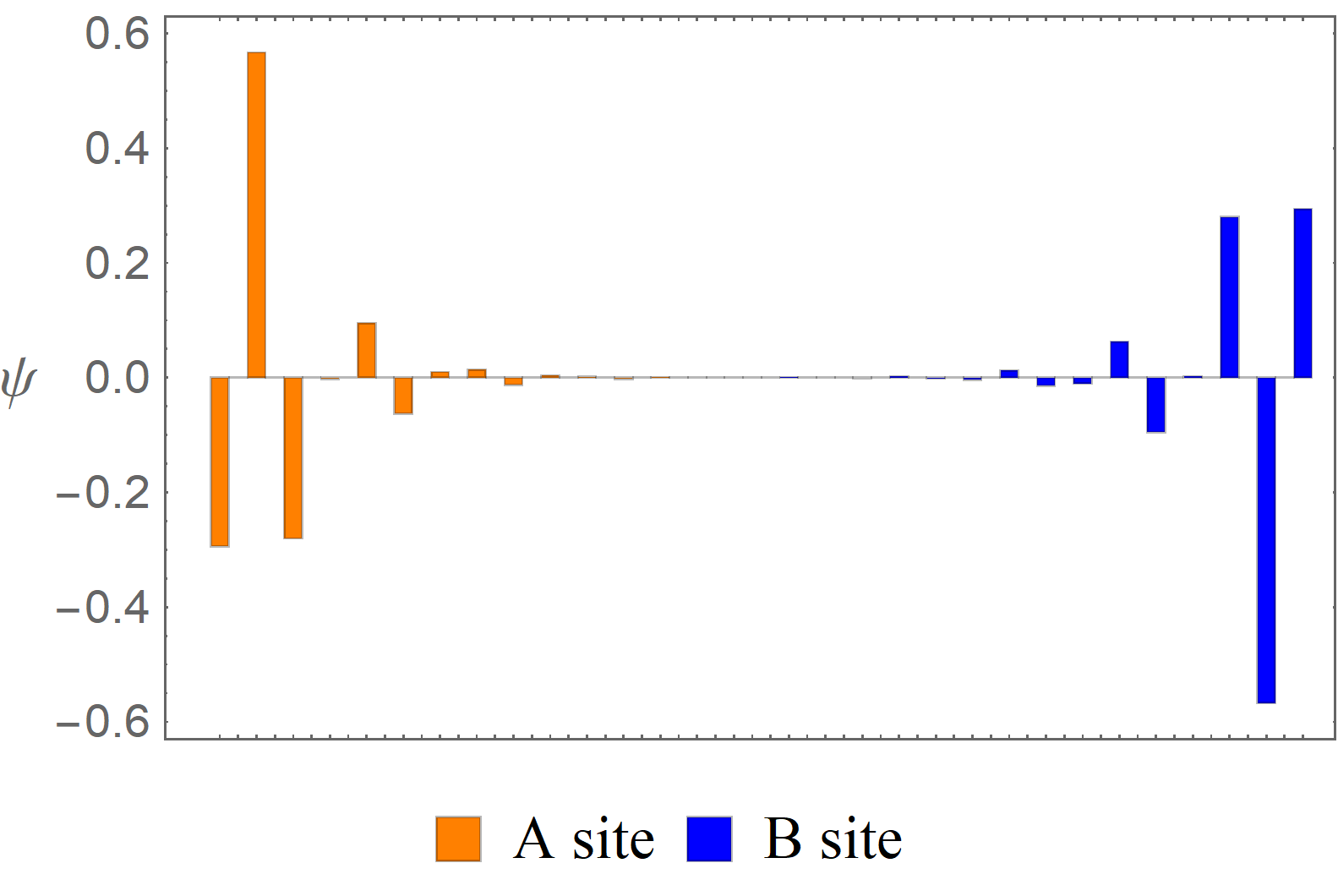}}\\
\subfloat[]{\includegraphics[width=0.30\textwidth]{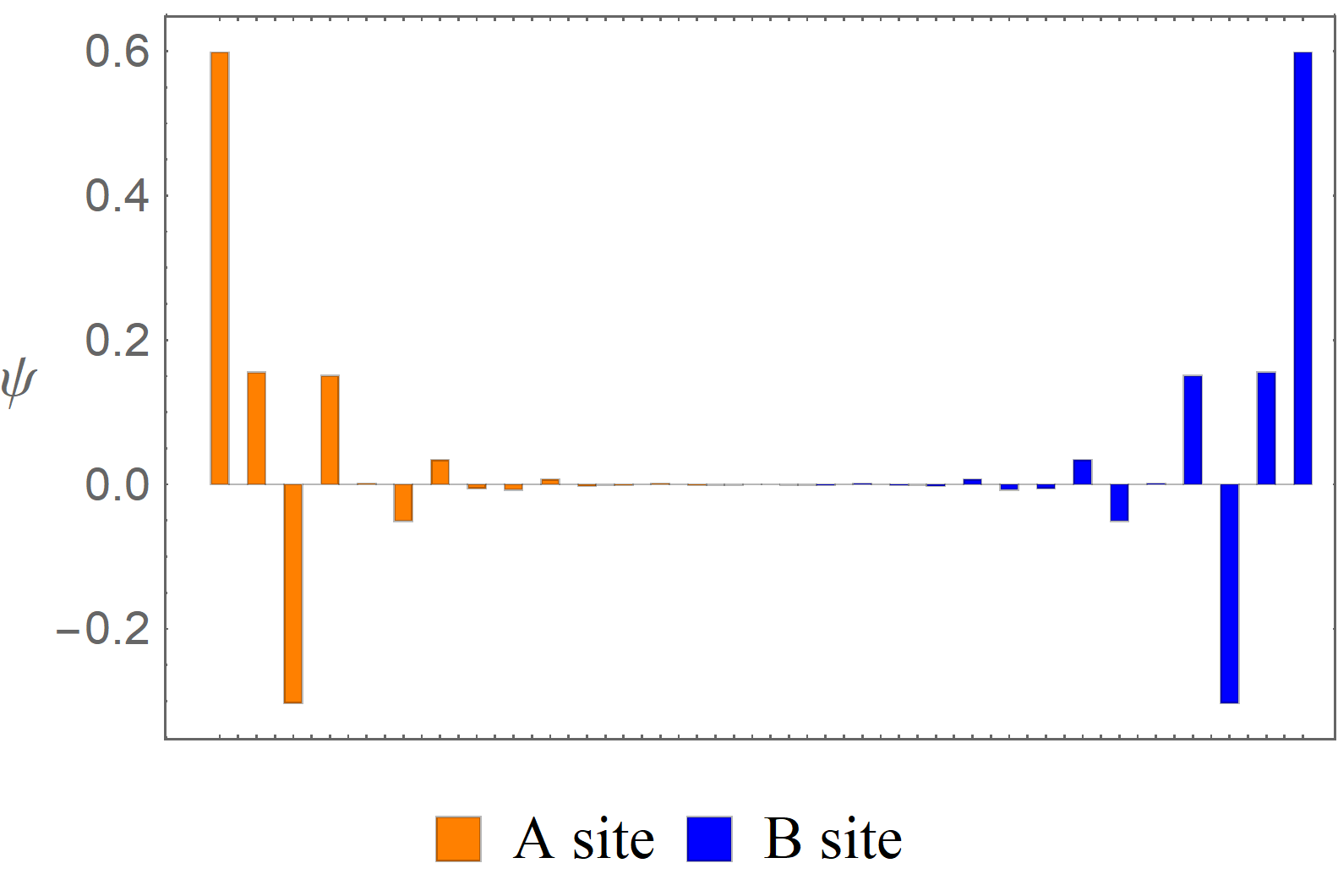}}\hskip 0.5cm
\subfloat[]{\includegraphics[width=0.30\textwidth]{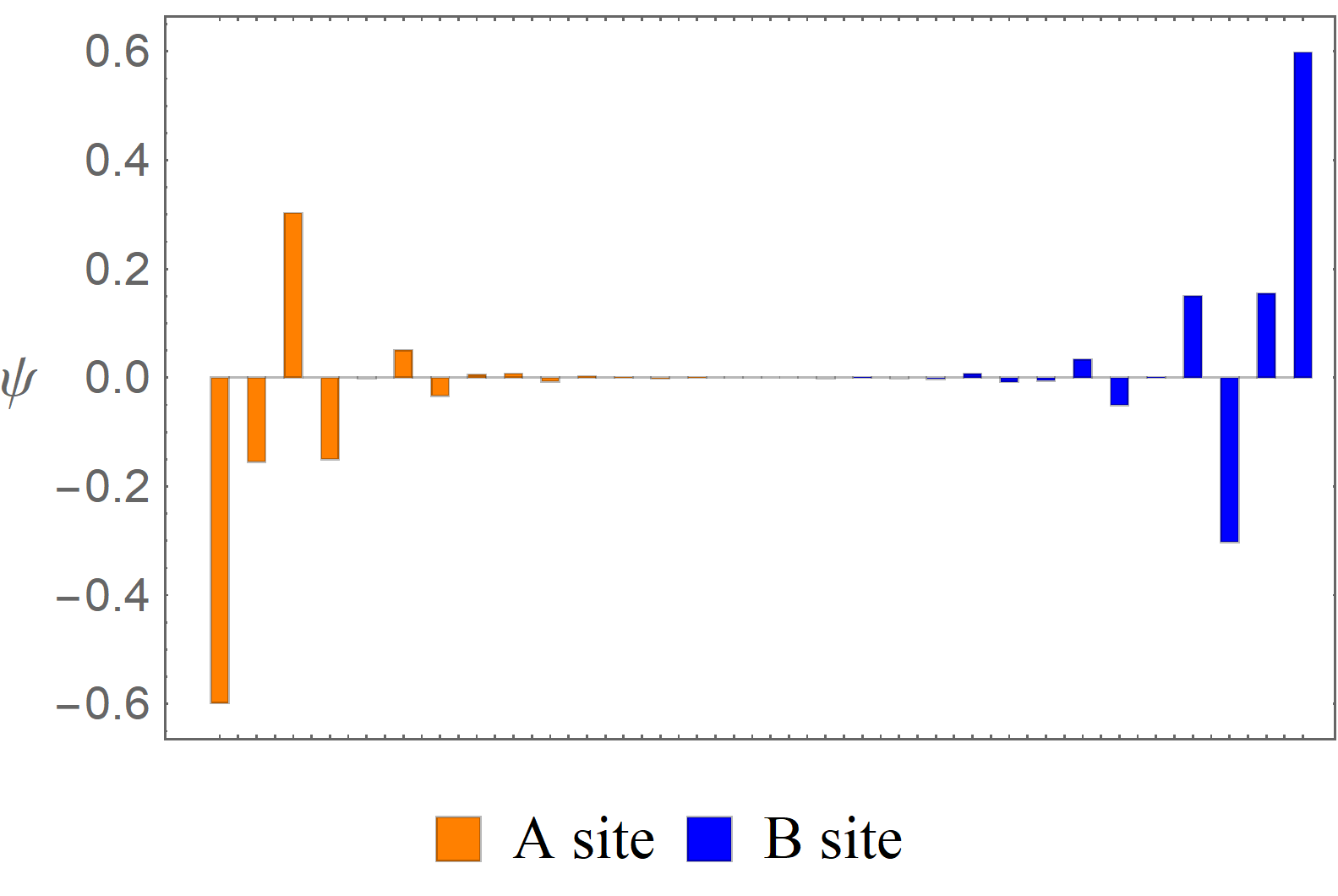}}\\
\subfloat[]{\includegraphics[width=0.32\textwidth]{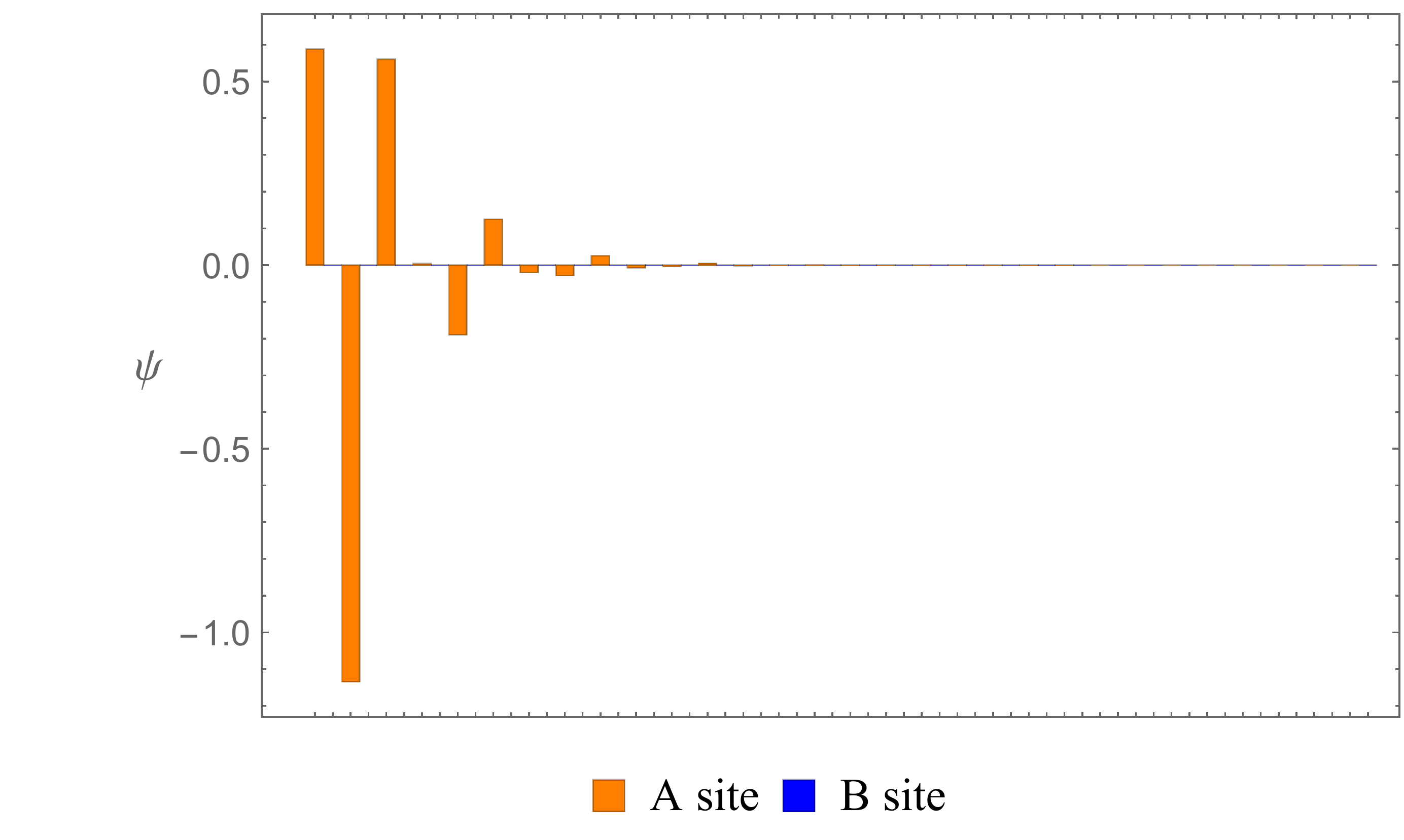}}\hskip 0.5cm
\subfloat[]{\includegraphics[width=0.32\textwidth]{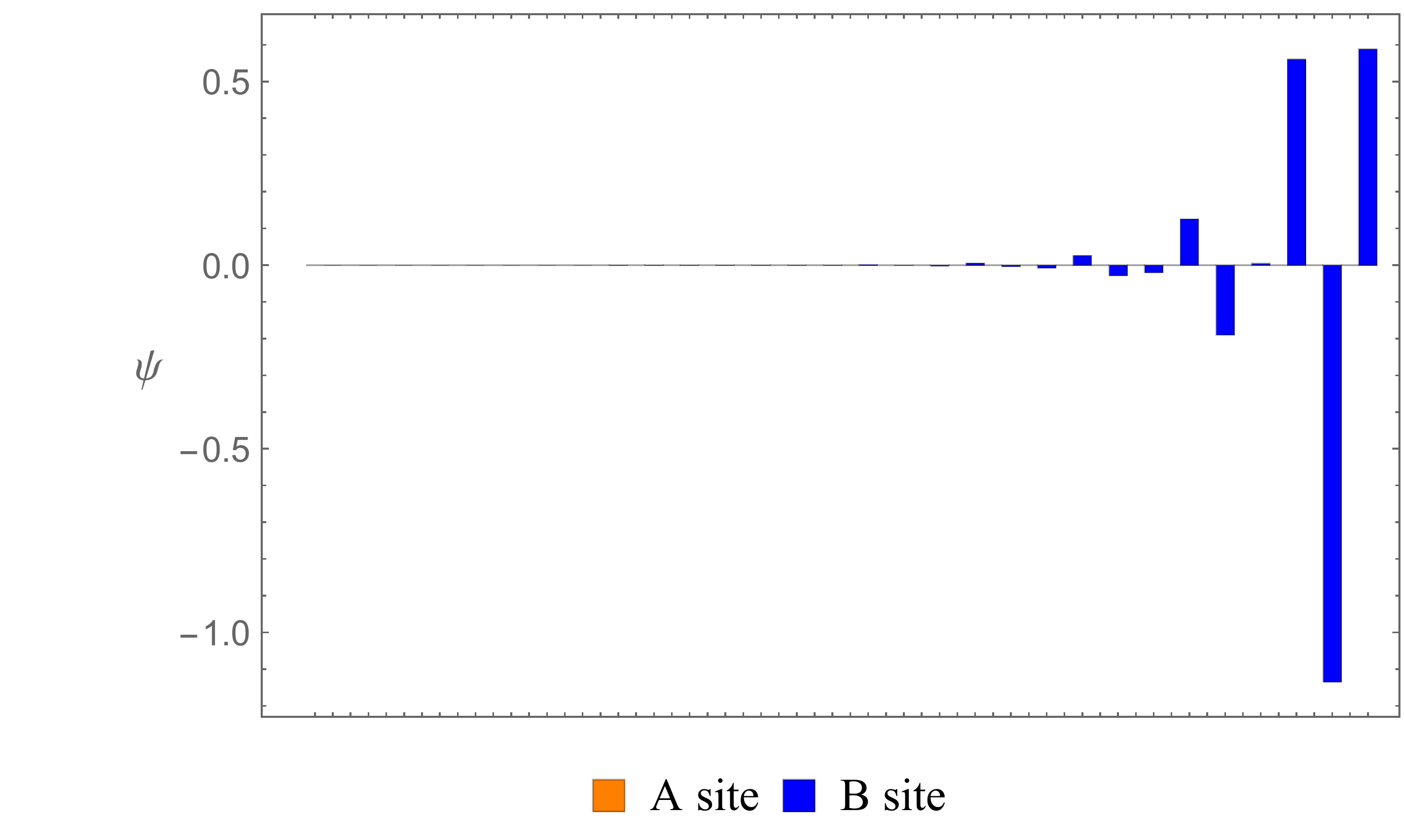}}\\
\subfloat[]{\includegraphics[width=0.32\textwidth]{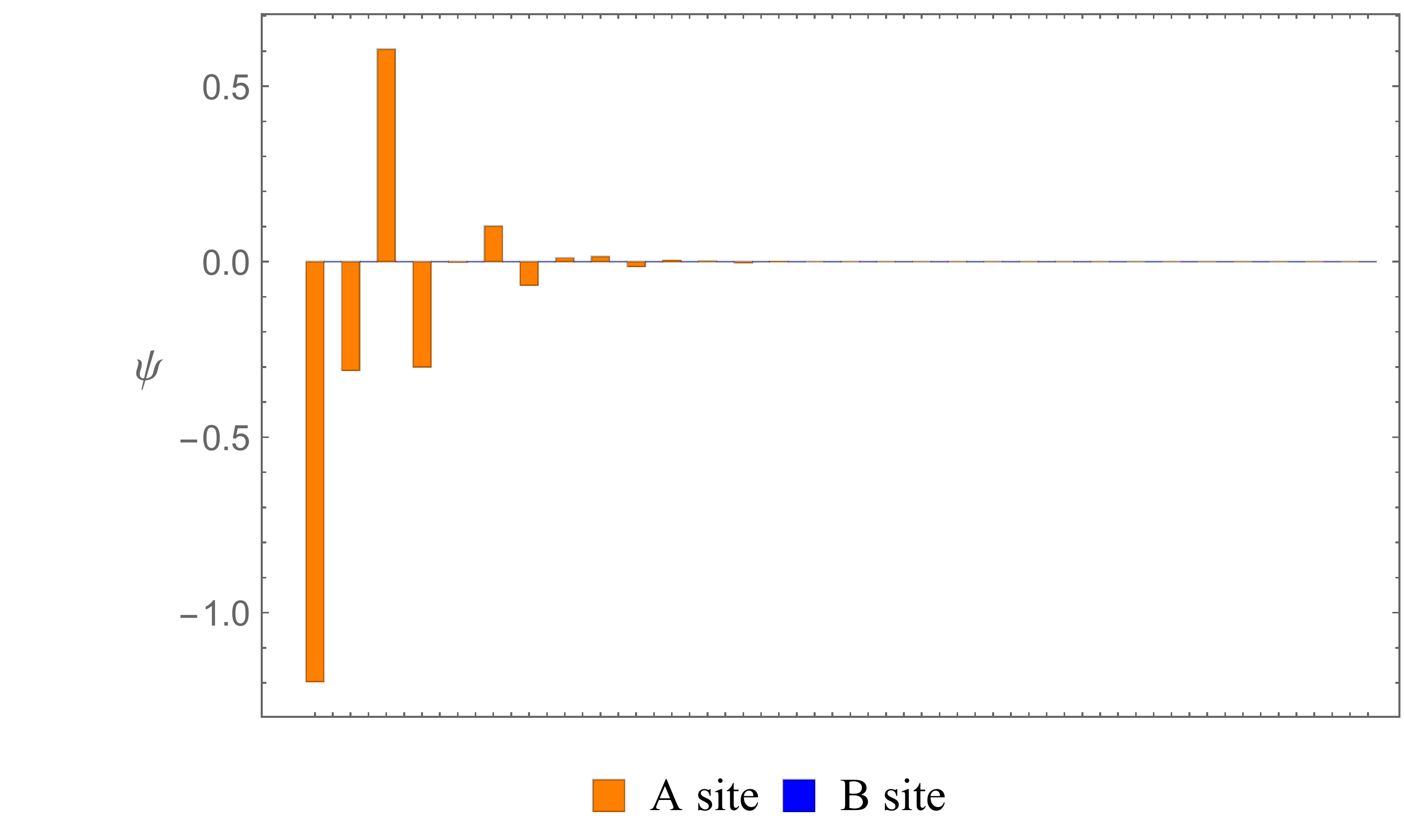}}\hskip 0.5cm
\subfloat[]{\includegraphics[width=0.32\textwidth]{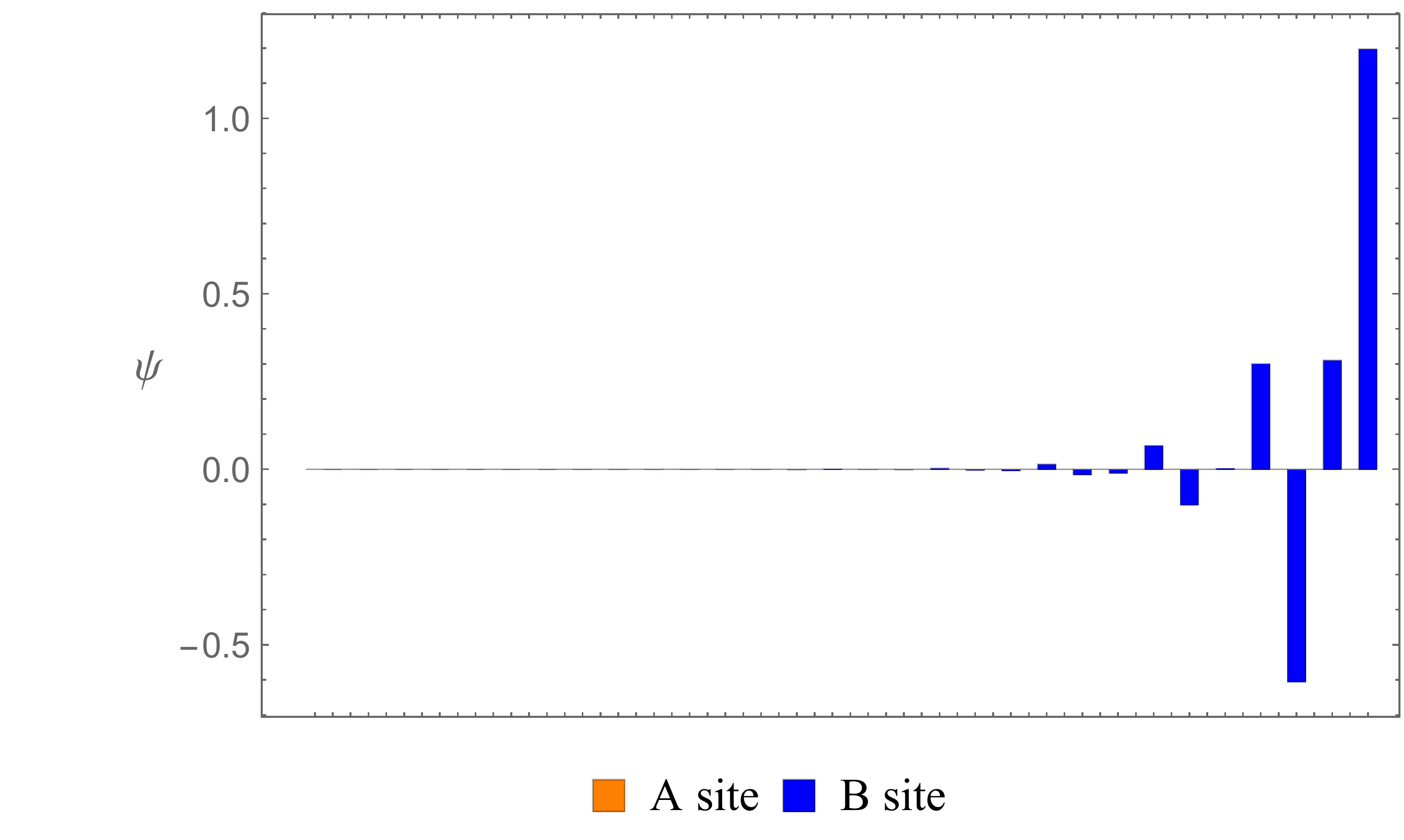}}
\caption{(a)The energy spectrum of the extended SSH model with $\n = 2$, where $(t_0,t_1,t_2)$ are $(5,10,15)$. (b)-(e) The wave functions of the four edge states with almost zero energy in the system.  (f),(g) By taking the difference and sum of the even and odd wave functions of the almost zero energy states in (b) and (c), we obtain a set of left and right edge states, respectively. (h),(i) By taking the difference and sum of the even and odd wave functions of the almost zero energy states in (d) and (e), we again obtain another set of left and right edge states, respectively.}\label{fig ext-ssh w2 even eigenvector}
\end{figure}

\begin{figure}[hbt!]
\centering
\subfloat[]{\includegraphics[width=0.50\textwidth]{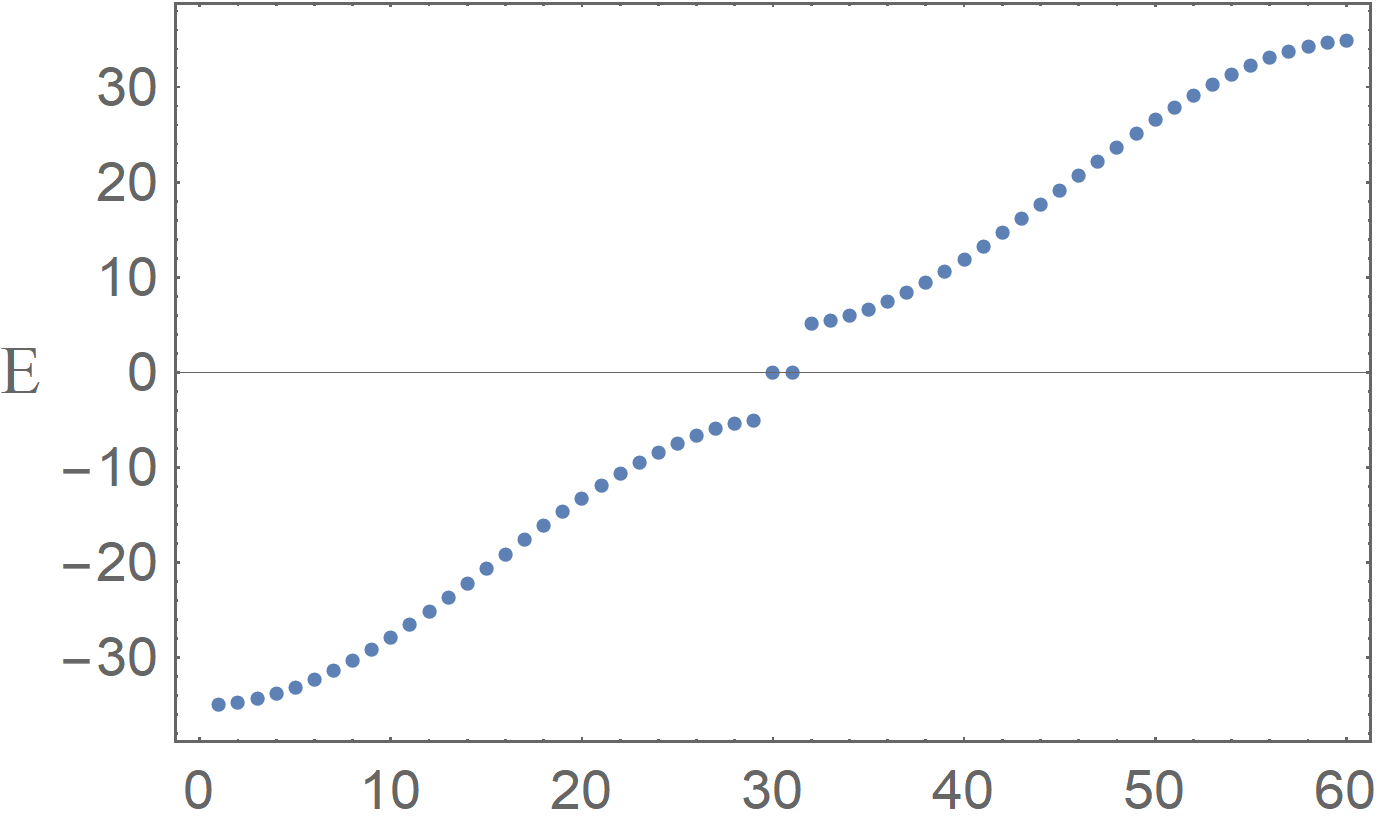}}\\
\subfloat[]{\includegraphics[width=0.30\textwidth]{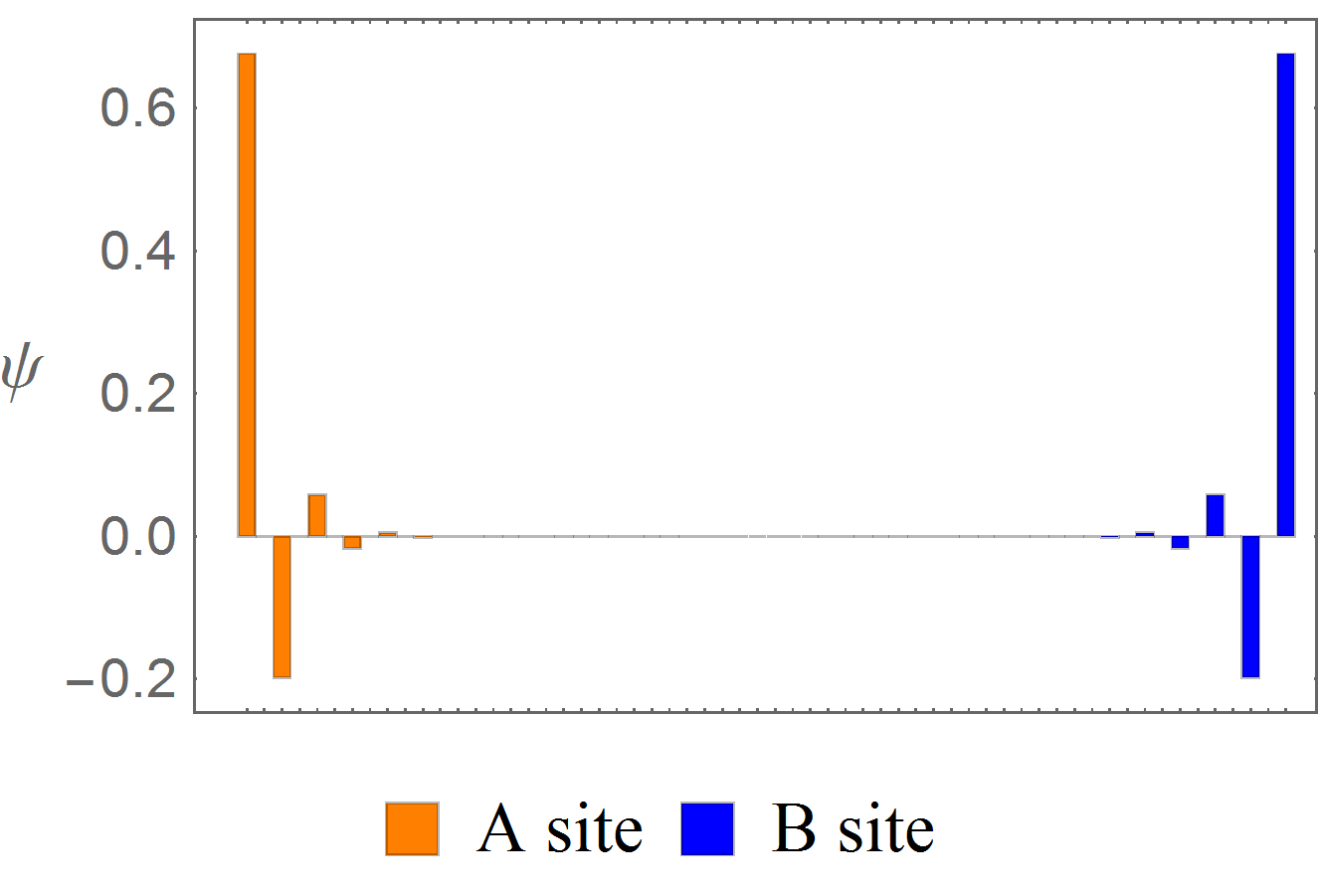}}\hskip 0.5cm
\subfloat[]{\includegraphics[width=0.30\textwidth]{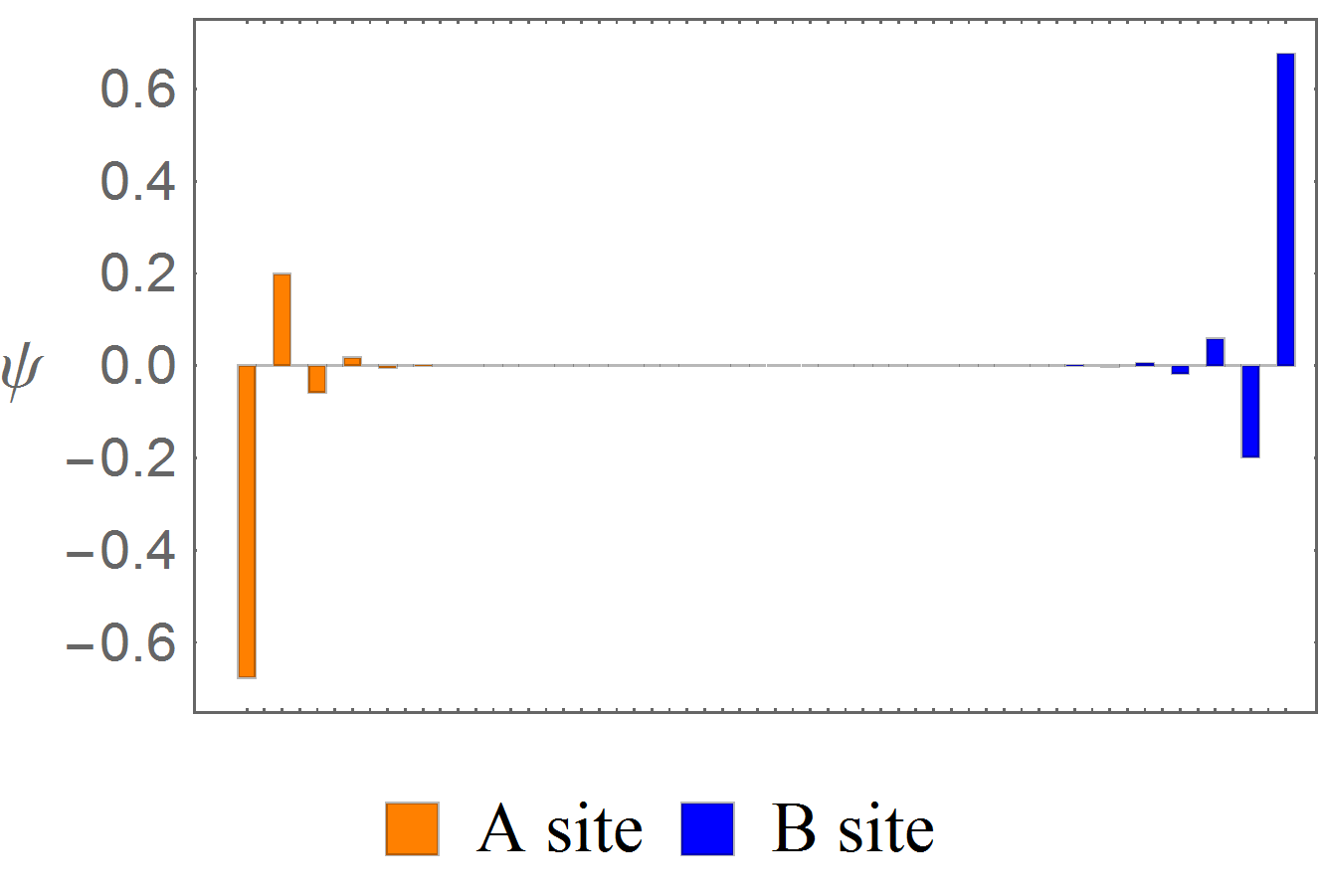}}
\caption{(a)The energy eigenvalue of the system with $\n = 1$, where $(t_0,t_1,t_2)$ are $(5,20,10)$. (b),(c) The wave functions of the two edge states with almost zero energy in the system.}\label{fig ext-ssh w1 even eigenvector}
\end{figure}

Again, the bulk-edge correspondence may be confirmed numerically by considering a finite chain of the extended SSH model. As an illustration, let's first consider the case that there are 60 (even) sites. In particular, we choose $(t_0,t_1,t_2)=(5,10,15)$ and $(t_0,t_1,t_2)=(5,20,10)$ so that the winding numbers are $\n=2$ and $\n=1$, respectively. The energy spectrum and the wave functions of the edge states for the two cases are shown in Fig.\ref{fig ext-ssh w2 even eigenvector} and Fig.\ref{fig ext-ssh w1 even eigenvector}.  Similar to the SSH model, we see that all the "zero-energy'' edge states involve mixing of the left and right edge states.  By taking the difference and sum of the "zero-energy'' edge states properly, we may again restore the left and right edge states. Again, edge states would decay exponentially away from the boundaries.  Since $s$ is a solution to the quadratic equation in Eq.~(\ref{quadratic equation of s}), it is generally complex. Similarly, we can find numerically the best fit to the base number of the exponential function: $-0.3333 + 0.4714i$ and $-0.3333 - 0.4714i$. For sure, they are well consistent with the roots given in Eq.~(\ref{solution of quadratic equation}). Therefore, the bulk-edge correspondence is uphold.

\begin{figure}[hbt!]
\centering
\subfloat[]{\includegraphics[width=0.34\textwidth]{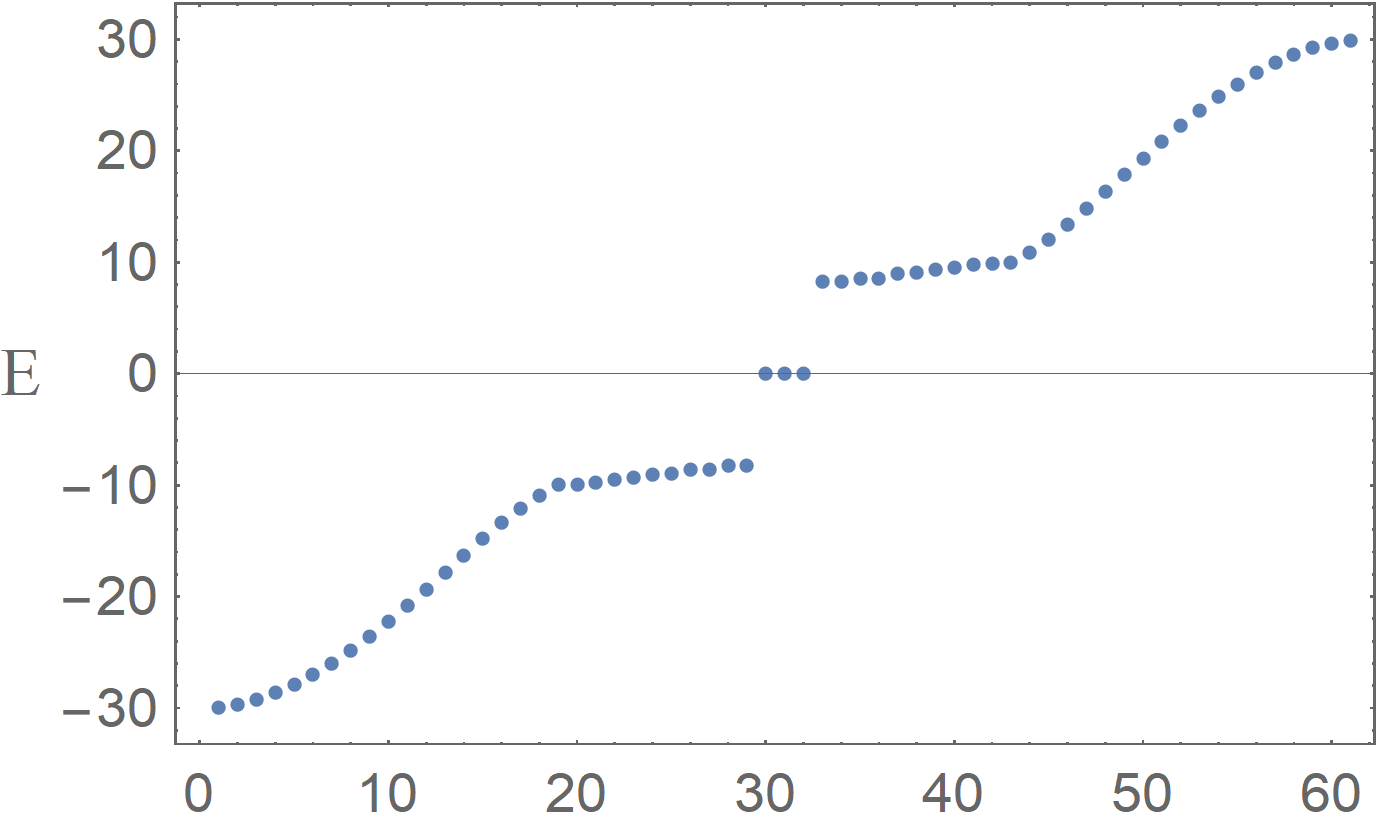}}\hskip 0.5cm
\subfloat[]{\includegraphics[width=0.30\textwidth]{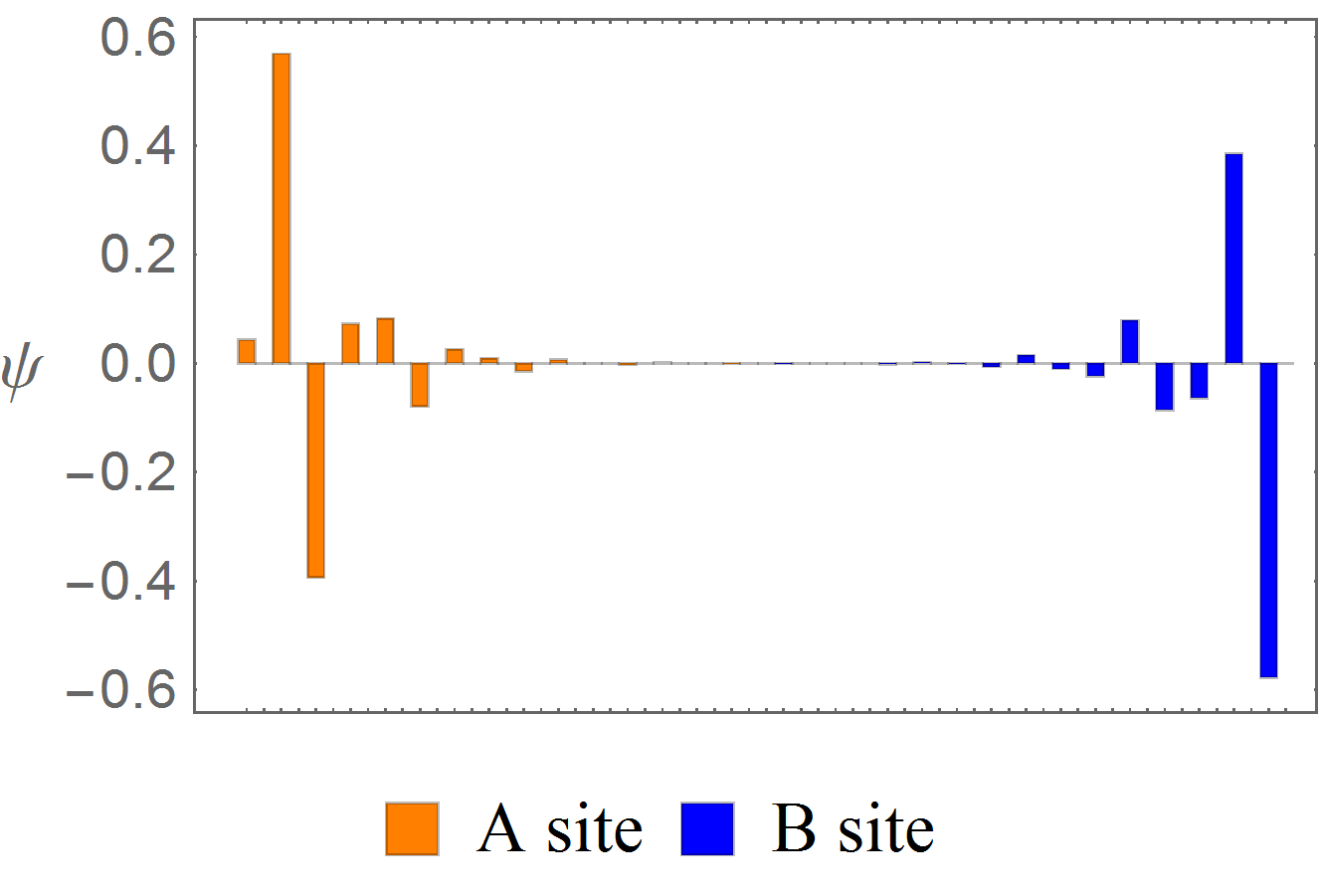}}\\
\subfloat[]{\includegraphics[width=0.30\textwidth]{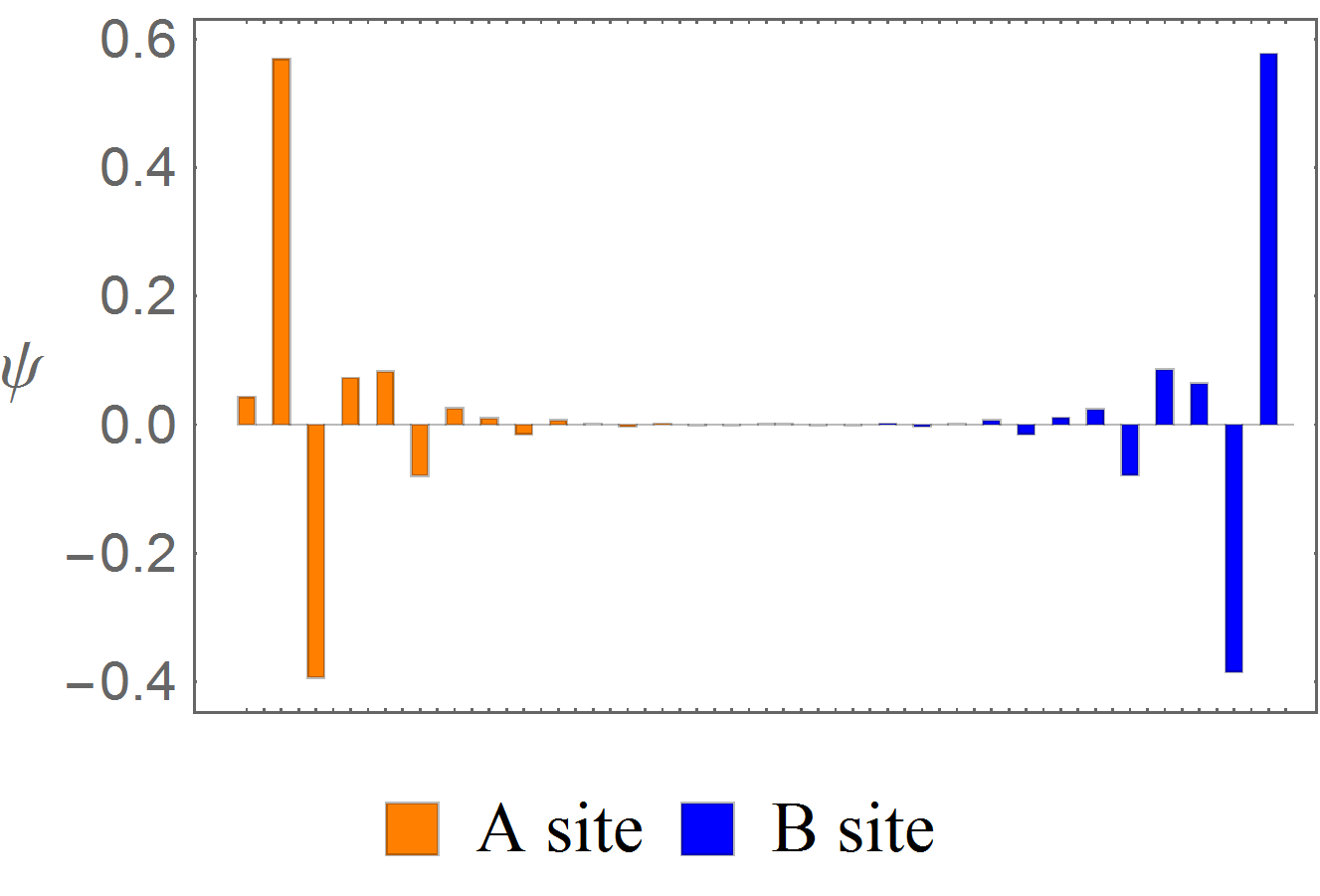}}\hskip 0.5cm
\subfloat[]{\includegraphics[width=0.30\textwidth]{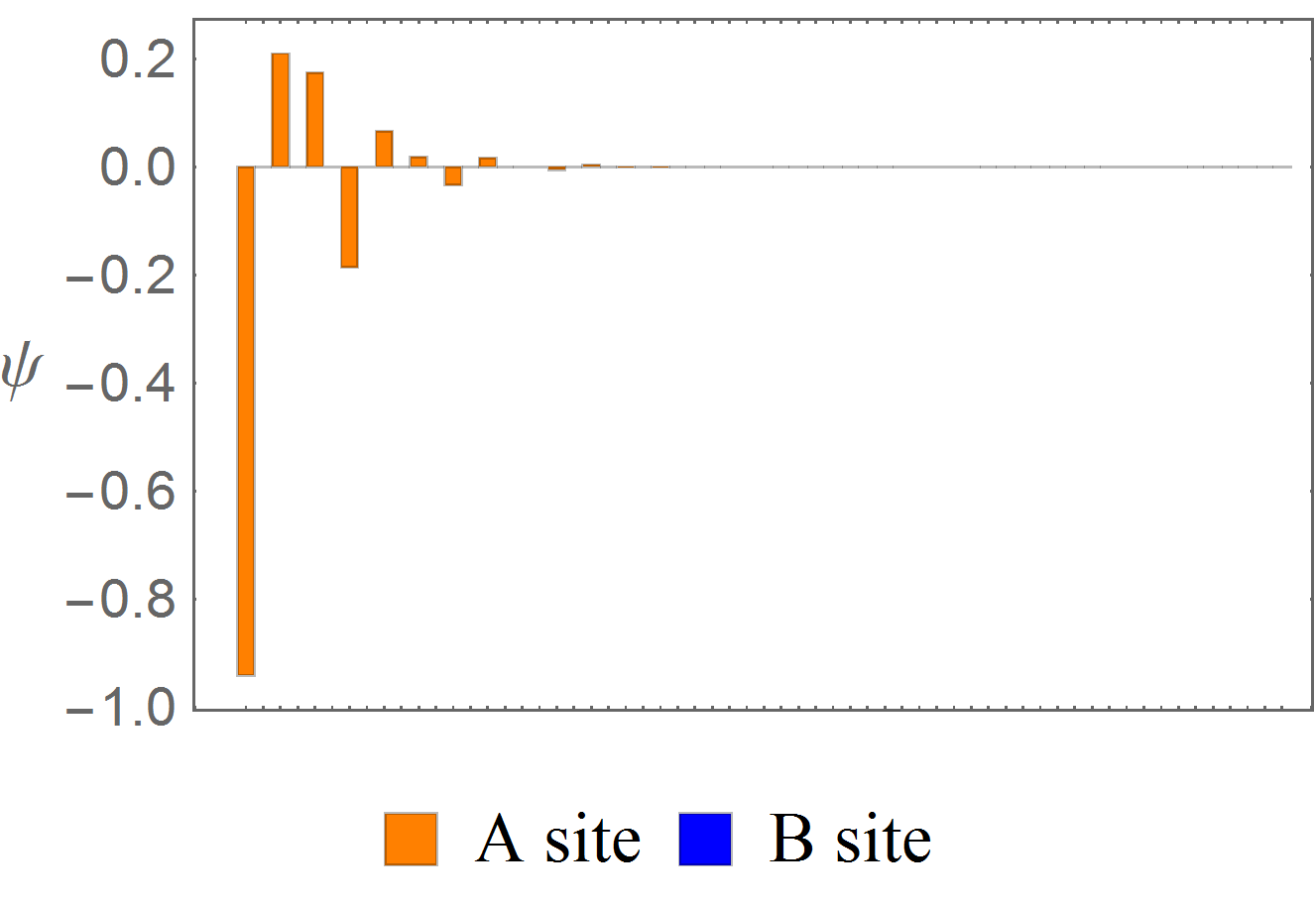}}
\caption{(a)The energy spectrum of the system for the case that $(t_0,t_1,t_2)=(5,10,15)$ and the total number of sites is 61. (b),(c) The wave functions of the edge states with almost zero energy in the system. (d)The edge state with exact zero energy in the system.}\label{fig ext-ssh w2 odd eigenvector}
\end{figure}

Next, let's consider the case that there are  61 (odd) sites. Similar to the SSH model, the unit cells consistent with the left and right boundaries are different. From Eq.~(\ref{ssh ext1 hamiltonian tilde}), we now have
\bea
&\;& \hskip -1.0cm \tilde{h}_{\mathrm{ext}1}(p) = \left(t_0 \e^{ip}+ t_1 + t_2 \e^{-ip} \right) = \e^{ip}h^*_{\mathrm{ext}1}(p).
\eea
As a result, $\tilde{\n}=1-\n$.  In other words, the winding numbers of the two edges are related by
\bea\label{left and right winding number}
&\;& \hskip -5.0cm \n_{\rm left}=1-\n_{\rm right}.
\eea
In the case $(t_0,t_1,t_2)=(5,10,15)$, we have $\n_{\rm left}=2$ and $\n_{\rm right}=-1$.  According to the bulk-edge correspondence, there should be two and one edge states on the left and right boundaries, respectively. Moreover, we expect one of the left edge states should have exact zero energy and is decoupled from all other edge states. The energy spectrum and the wave functions of the three edge states shown in Fig.~\ref{fig ext-ssh w2 odd eigenvector} confirm all these predictions.

\begin{figure}[hbt!]
\centering
\subfloat[]{\includegraphics[width=0.34\textwidth]{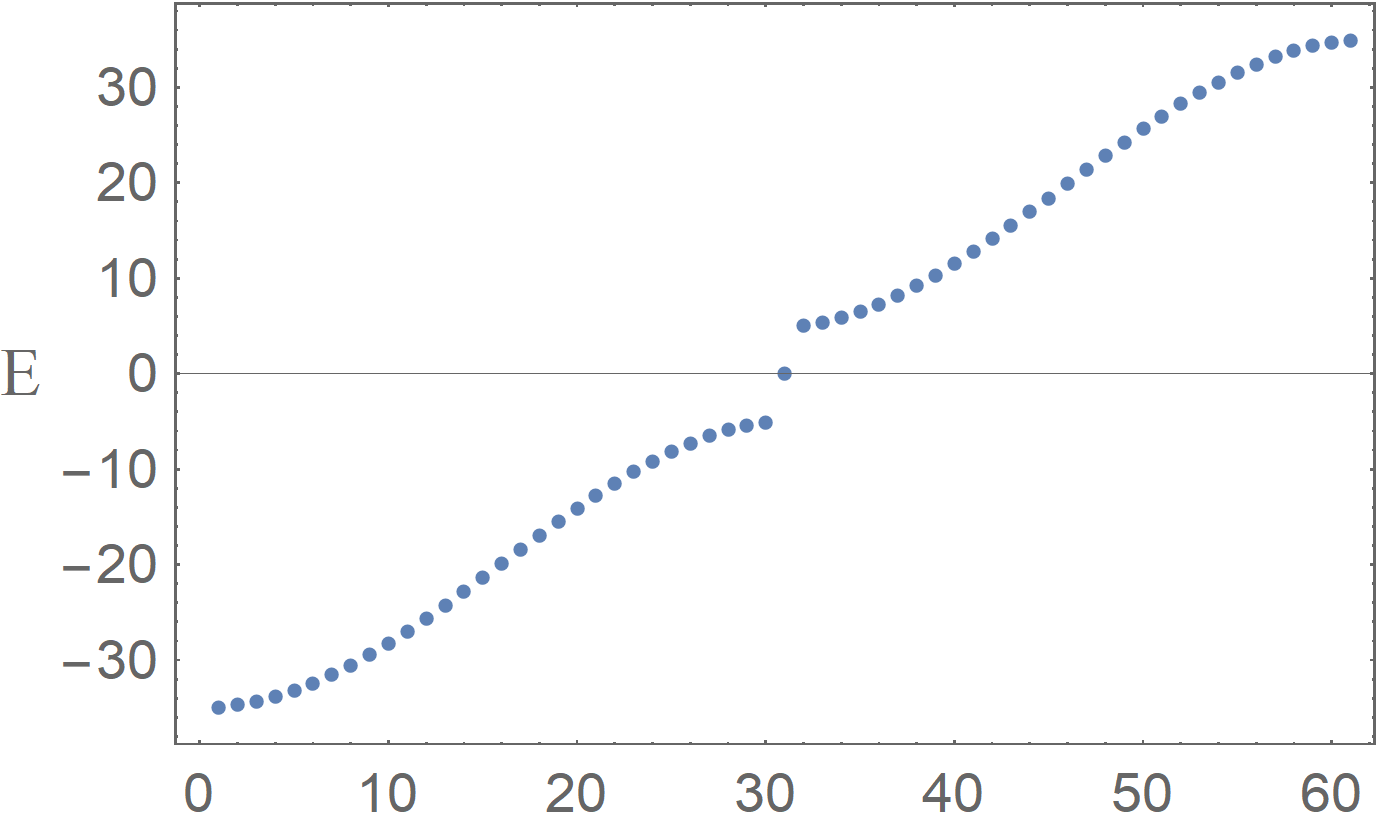}}\hskip 0.5cm
\subfloat[]{\includegraphics[width=0.30\textwidth]{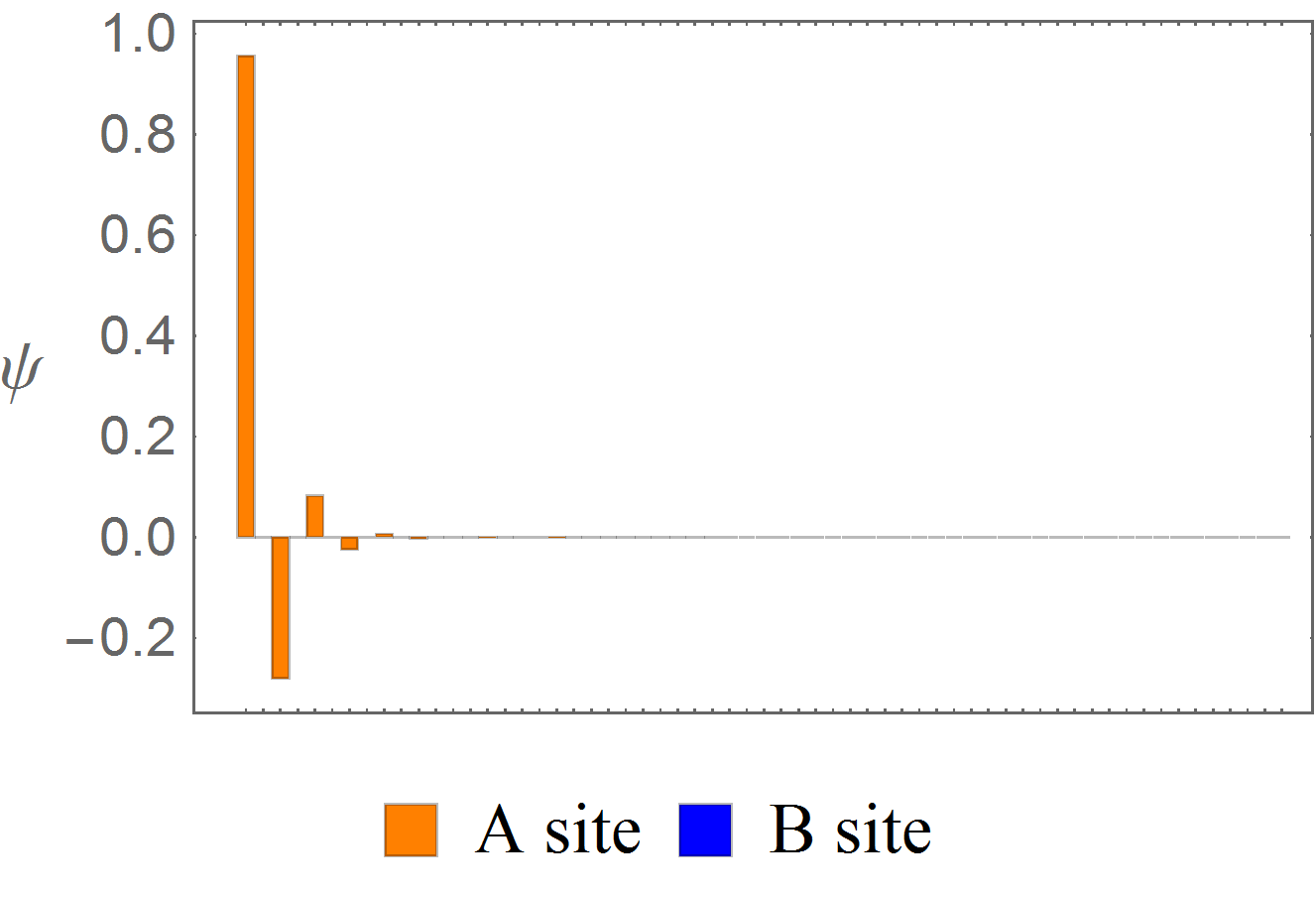}}
\caption{(a)The energy spectrum of the system for the case that $(t_0,t_1,t_2)=(5,20,10)$ and the total number of sites is 61. (b) The edge state wave function with zero energy in the system.}\label{fig ext-ssh w1 odd eigenvector}
\end{figure}

For $(t_0,t_1,t_2)=(10,20,5)$, we have $\n_{\rm left}=1$ and $\n_{\rm right}=0$.  According to the bulk-edge correspondence, there should be only one edge states on the left boundary. This again is confirmed by the energy spectrum and the wave functions of the edge states shown in Fig.\ref{fig ext-ssh w1 odd eigenvector}.  Similarly, we expect the left edge states should have exact zero energy and is decoupled from the right boundary.

\begin{figure}[hbt!]
\centering
\includegraphics[scale = 0.5]{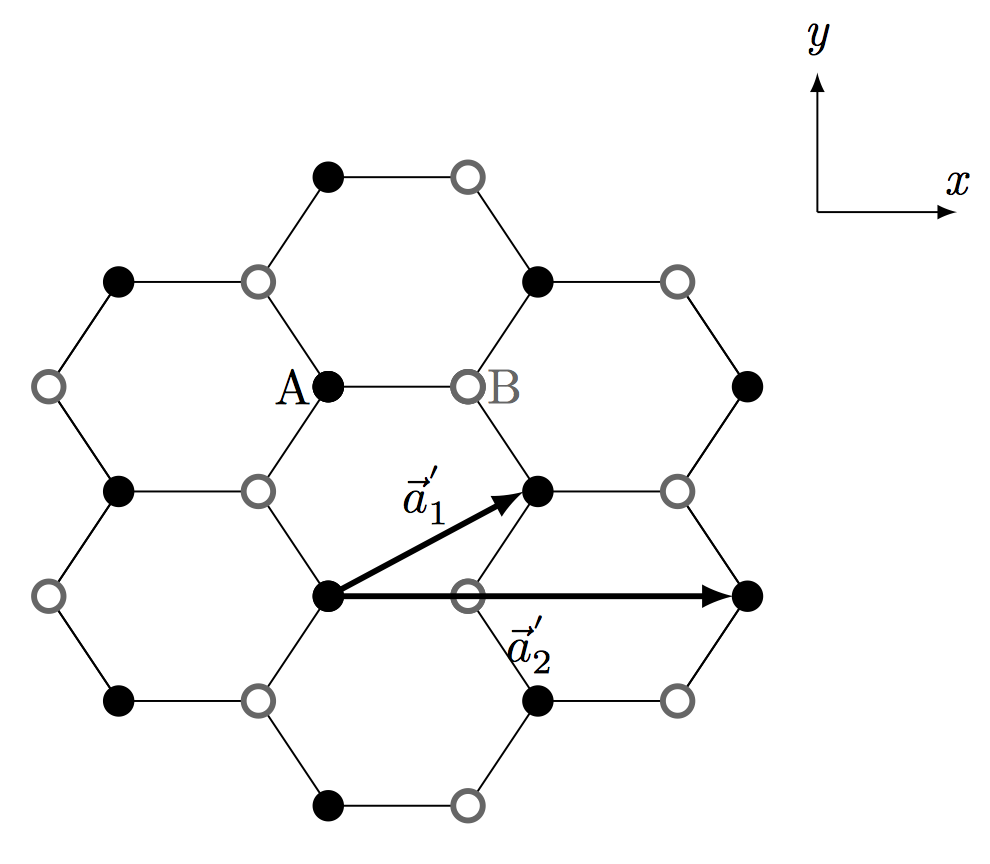}
\caption{Schematic diagram of graphene adopted from Ref. \cite{graphene}.}\label{graphene fig}
\end{figure}

We may use the results obtained in the SSH and extended SSH models to understand when and how the edge states in a carbon nanotube (CNT) with various edges would appear.  It is well-known that the graphene Hamiltonian may be cast in the following form \cite{graphene}
\bea
&\;& \hskip -1.2cm H =  \sum_{j_1,j_2} t \left\{ A^\dagger_{n_1,n_2} +  A^\dagger_{j_1+1j_2} +  A^{\dag}_{j_1-1,j_2+1} \right\} B_{j_1,j_2}  + {\rm h.c.} .
\eea
The location of an $A$ site is described by
\bea
\vec{r}_A = j_1 \vec{a}'_1+ j_2 \vec{a}'_2,
\eea
where we choose
\bea
\vec{a}'_1 = a \Big( \frac{\sqrt{3}}{2} , \frac{1}{2} \Big),\; \vec{a}'_2  = a \Big( \sqrt{3} , 0 \Big),
\eea
for convenience. Note that $a = \sqrt{3} a_{\rm cc}$, with $a_{\rm cc} = 1.42 \overset{\circ}{A}$, the carbon-carbon distance in graphene. In this convention, it would be straight forward to reduce the above Hamiltonian to those of the zigzag and armchair CNT's. See Fig.~\ref{graphene fig}.

First, let's impose the periodic boundary condition along the $\vec{a}'_1$ direction in graphene to obtain the Hamiltonian of the zigzag CNT:
\bea
&\;& \hskip -1.5cm H_{\rm zig} = \sum_{j_2} t \Big\{ (1+ e^{i p_1} )A^{\dag}_{j_2}+e^{-i p_1}  A^{\dag}_{j_2+1} \Big\} B_{j_2}  + {\rm h.c.}.
\label{zigzag}
\eea
Here, $p_1 = \frac{2 \pi k_1 }{a N_1}$ with $N_1$ the layer number along $\vec{a}'_1$ and $k_1 = 0,2,...,N_1-1$.  Since $p_1$ is a good quantum number, it is obvious that the above Hamiltonian is effectively one dimensional and is closely related to that of the SSH model.  It is transparent to see that the zigzag edge on the upper left boundary gives rise to the boundary conditions $B_0 = 0$.  By adding an extra layer of B sites, the upper left boundary becomes the zigzag beard edge, which leads to the boundary conditions $A_0 =0$. Meanwhile, the zigzag and zigzag beard edges on the lower right boundary give rise to the boundary condition $ A_{N_2+1}= 0$ and $B_{N_2+1} = 0$, respectively.  It is quite obvious that the there are two transition points in $p_1$ that separate the topological and trivial phases. It may be determined by the condition that
\bea
&\;& \hskip -6.5cm |1+ e^{i p_{\rm c}} | =1,
\eea
and thus $p_{\rm c} = 2\pi/3, 4\pi/3$.

\begin{figure}[hbt!]
\centering
\includegraphics[scale = 0.5]{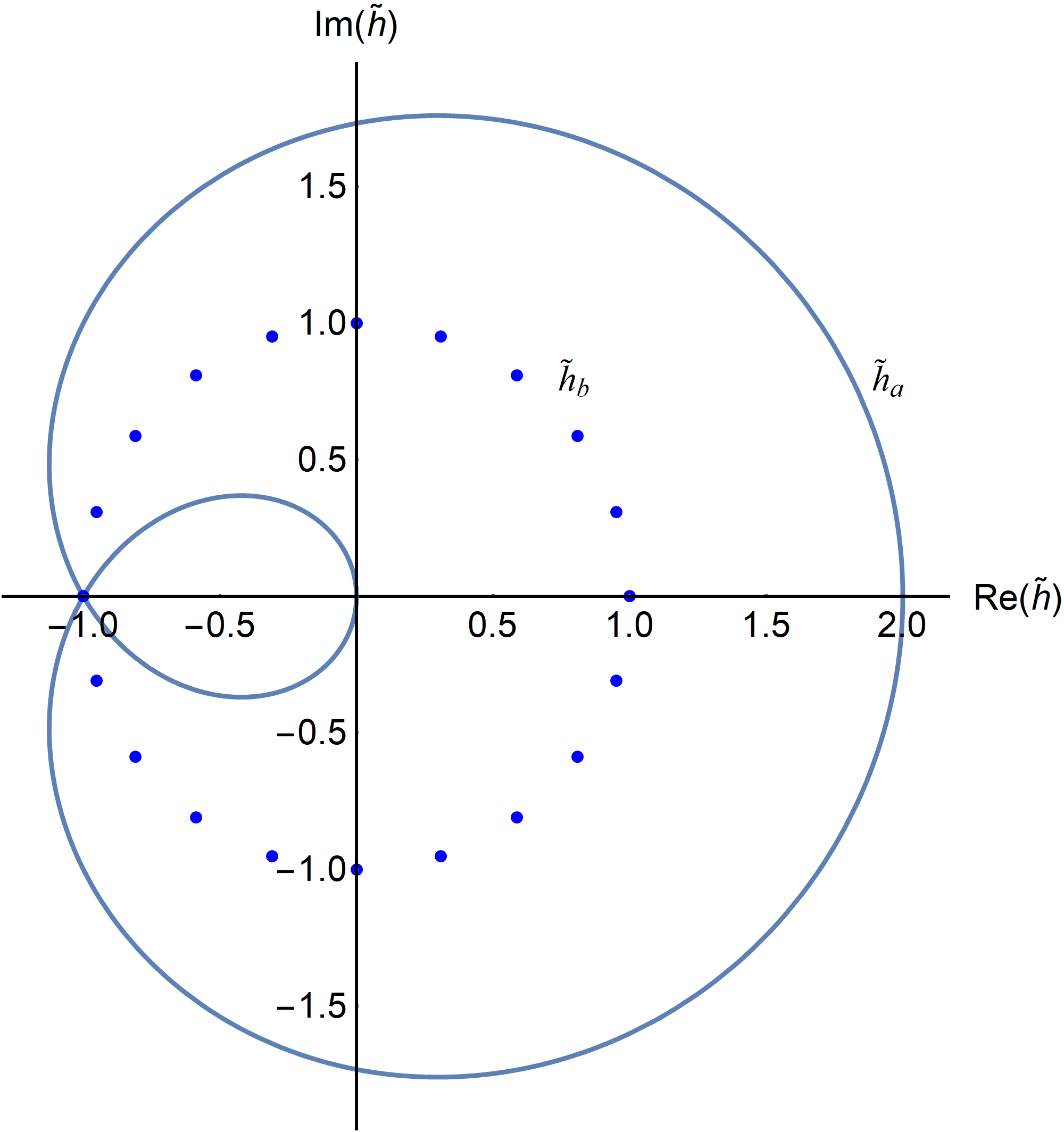}
\caption{Trajectory of $\tilde{h}_a$ and $\tilde{h}_b$. The winding number is always 1 on the armchair beard edge except for $p_2 =0$.} \label{armchair-beard CNT}
\end{figure}

Next, by imposing the periodic boundary condition along the $\vec{a}'_2$ direction, we obtain the Hamiltonian of the armchair CNT:
\bea
&\;& \hskip -1.5cmH_{\rm arm} = \sum_{j_1} t \Big\{ A^{\dag}_{j_1}+A^{\dag}_{j_1+1,}+ e^{ip_2} A^{\dag}_{j_1-1} \Big\} B_{j_1} + {\rm h.c.}.
\label{armchair}
\eea
Here, $p_2 = \frac{2 \pi k_2 }{\sqrt{3}aN_2}$ with $N_2$ the layer number along $\vec{a}'_2$ and $k_2 = 0,2,...,N_2-1$.  Similarly, it is obvious that the above Hamiltonian is effectively 1D and resembles that of the type 2 extended SSH model. Again, it may be seen that on the lower boundary the armchair and armchair beard edges lead to the boundary condition $A_{0} = 0, B_{0} = 0$ and $B_{0} = 0, B_{-1} = 0$, respectively.  Of course, there is also a similar correspondence between the edges and boundary conditions on the upper boundary.  Here,  the $h$ in the Bloch Hamiltonian associated with the armchair edge is given by
\bea
h(p_1) = t \left(1 + \e^{ip_1} + \e^{-i p_1- i p_2} \right),
\eea
where $p_2$ is a good quantum number and should be considered a constant. When $p_1$ goes over the Brillouin zone, $h(p_1)$ trace out a straight line. Since the corresponding winding number is zero, a CNT with the armchair edge is usually known to be in the trivial phase and there would not be any edge state.  On the other hand, the Bloch Hamiltonian associated with the armchair beard edge is given by
\bea
\tilde{h}(p_1) = t \left(\e^{ip_1} + \e^{2ip_1} + \e^{- i p_2} \right).
\eea
Define $\tilde{h}_a = \e^{ip_1} + \e^{2ip_1}$ and $\tilde{h}_b = -\e^{- i p_2}$. From Fig.~\ref{armchair-beard CNT}, we see that except for the point $p_2 =0$, the winding number is always 1 and there would be edge states on the corresponding boundary.
The main lesson we learn here is that the existence of edge states is closely related the details of the boundary condition.  Since CNT may be synthesized in laboratories, it would be interesting to carry out relevant measurements to check the above prediction.

Before ending this section, we would like to mention that further generalization may be made by including next next to nearest neighbor hoping amplitudes and so on such that the corresponding Bloch Hamiltonian takes the form
\begin{equation}
\mathcal{H} (p)=
\left( \begin{array}{cc}
   0 &h^* (p)\\
  h(p) & 0
   \end{array} \right),
\end{equation}
with
\bea
 h(p)=\underset{m=-m_-}{\overset{m_+}{\sum }}t_j e^{imp}.
\eea
The value of the winding number $\n$ would now lie in the range $[-m_-, m_+]$ \cite{Chen-Chiou}.

\subsection{III. Connection between the winding number and the Chern number}

It is well-known that we may introduce an on-site energy term in the SSH model to achieve the Rice-Mele model \cite{Rice-Mele}. Its Bloch Hamiltonian is given by
\be\label{RM}
\HH_{\rm RM}=\left(
\begin{matrix}
m_0 & t_0+t_1 \e^{-ip} \cr
t_0+t_1 \e^{ip} & -m_0 \cr
\end{matrix}
\right).
\ee
When we consider a right semi-infinite chain of such a model, the recurrence relation and boundary condition are given by
\bea
&\;& \hskip -3.1cm \left(E-m_0 \right)A_j + \left(t_0 B_j +t_1 B_{j-1} \right) =0; \cr
&\;& \hskip -3.1cm \left(E+m_0 \right)B_j +\left(t_0 A_j +t_1 A_{j+1} \right) =0, \cr
&\;& \hskip -3.1cm B_0 =0.
\eea
Thus, an edge state exists only if $t_1>t_0$.  Because of the existence of on-site energy, the energy of the edge state is now shifted to
\bea
&\;& \hskip -6.1cm E = m_0,
\eea
and the wave function of the edge state is given by
\bea
&\;& \hskip -3.1cm A_j = A_1\left(-t_0/t_1 \right)^{j-1},\;  B_j =0,
\eea
with $j\ge 1$. Similarly, if we consider a right semi-infinite chain starting with the $B_0$ site instead, an edge state with $E=-m_0$ exists only if $t_1<t_0$.

\begin{figure}[hbt!]
\centering
\subfloat[]{\includegraphics[width=0.30\textwidth]{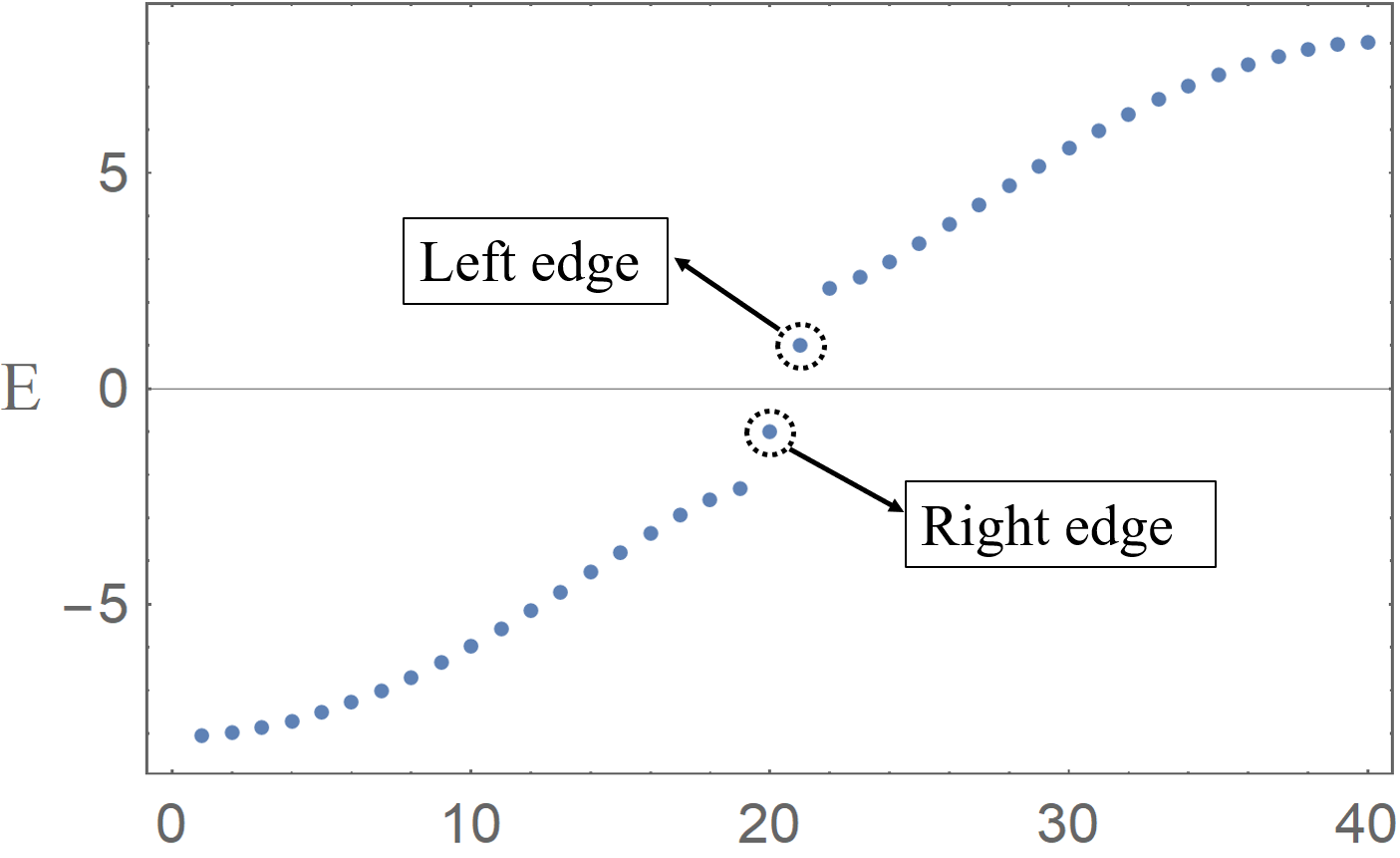}}\hskip 0.5cm
\subfloat[]{\includegraphics[width=0.30\textwidth]{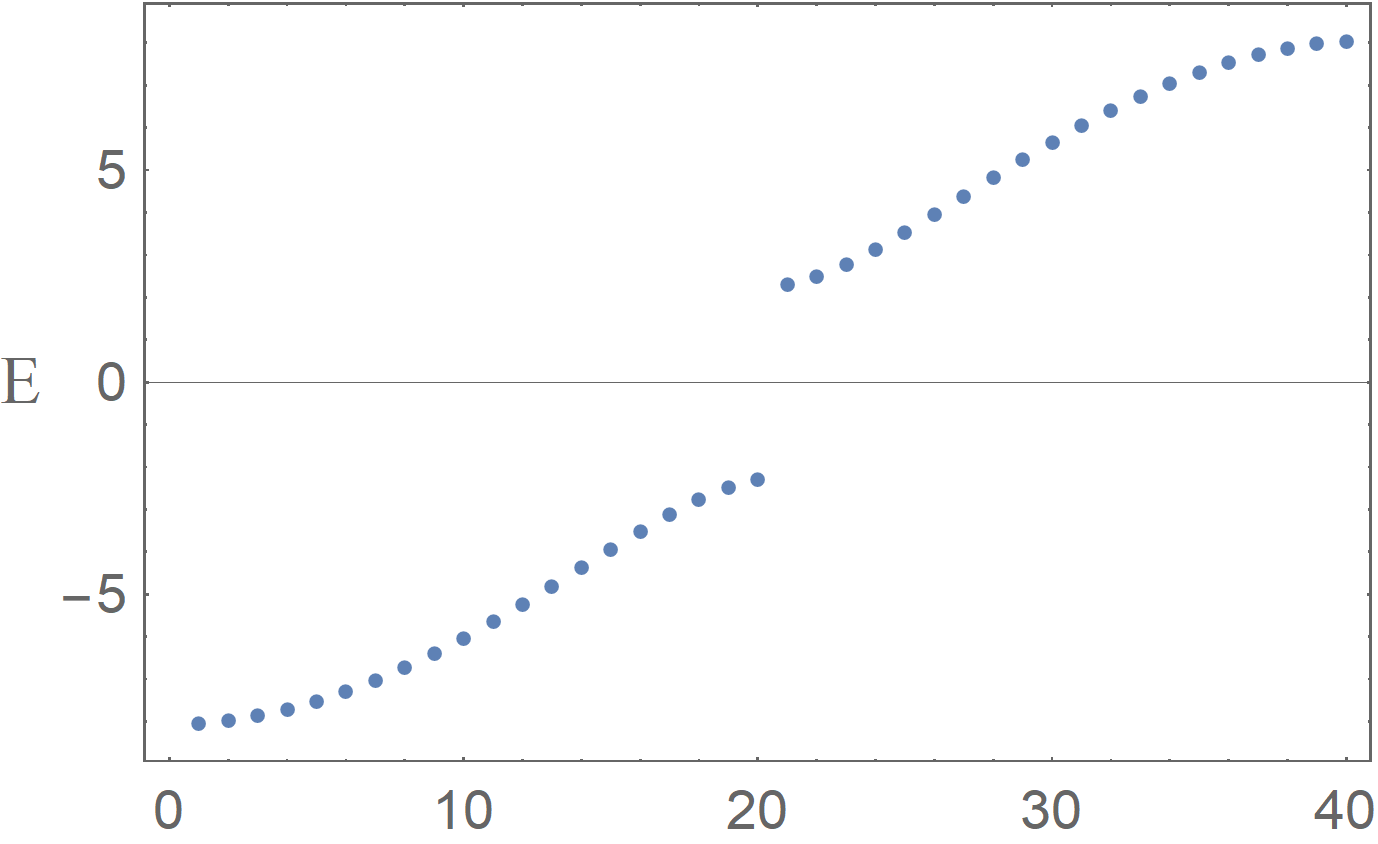}}\\
\subfloat[]{\includegraphics[width=0.30\textwidth]{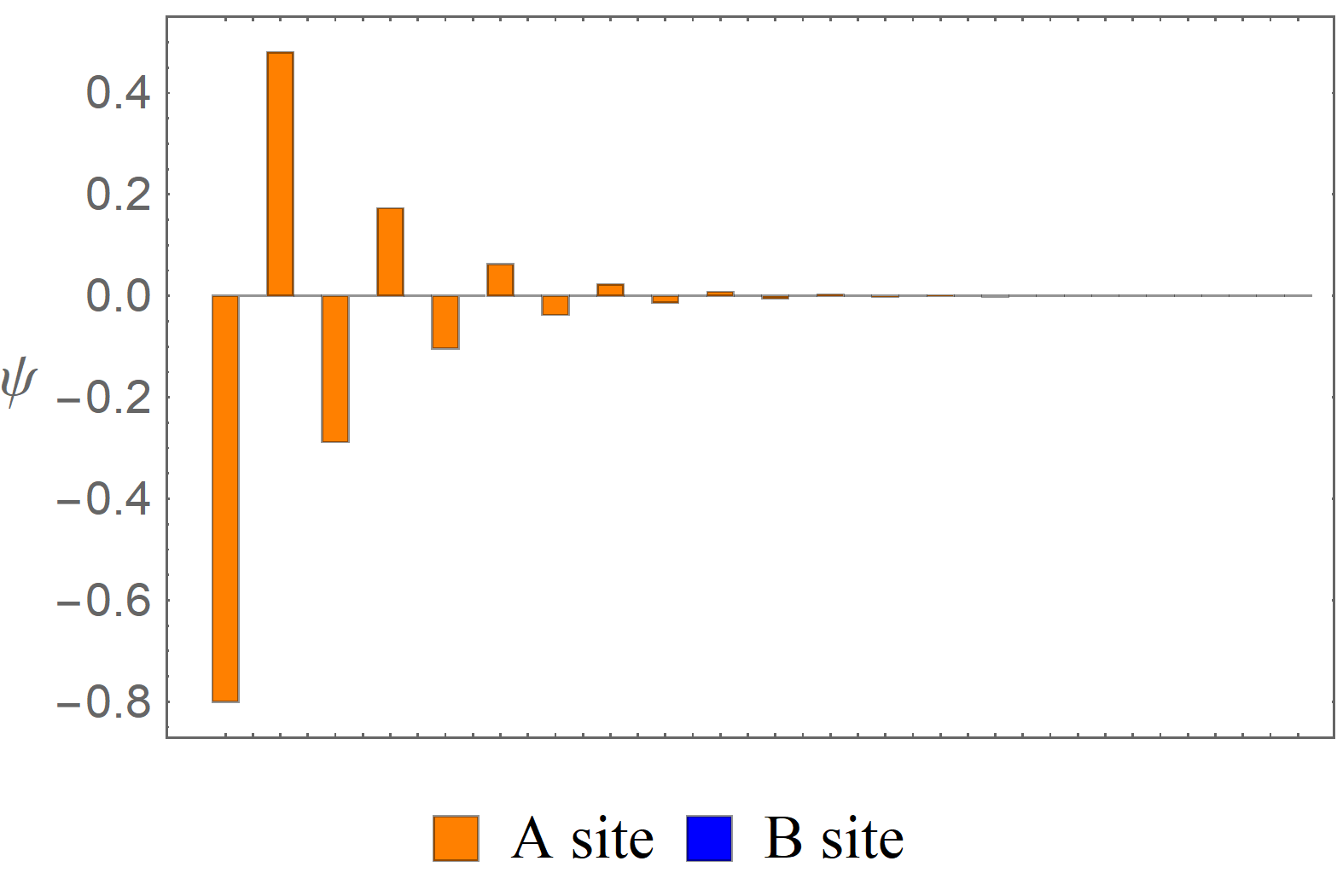}}\hskip 0.5cm
\subfloat[]{\includegraphics[width=0.30\textwidth]{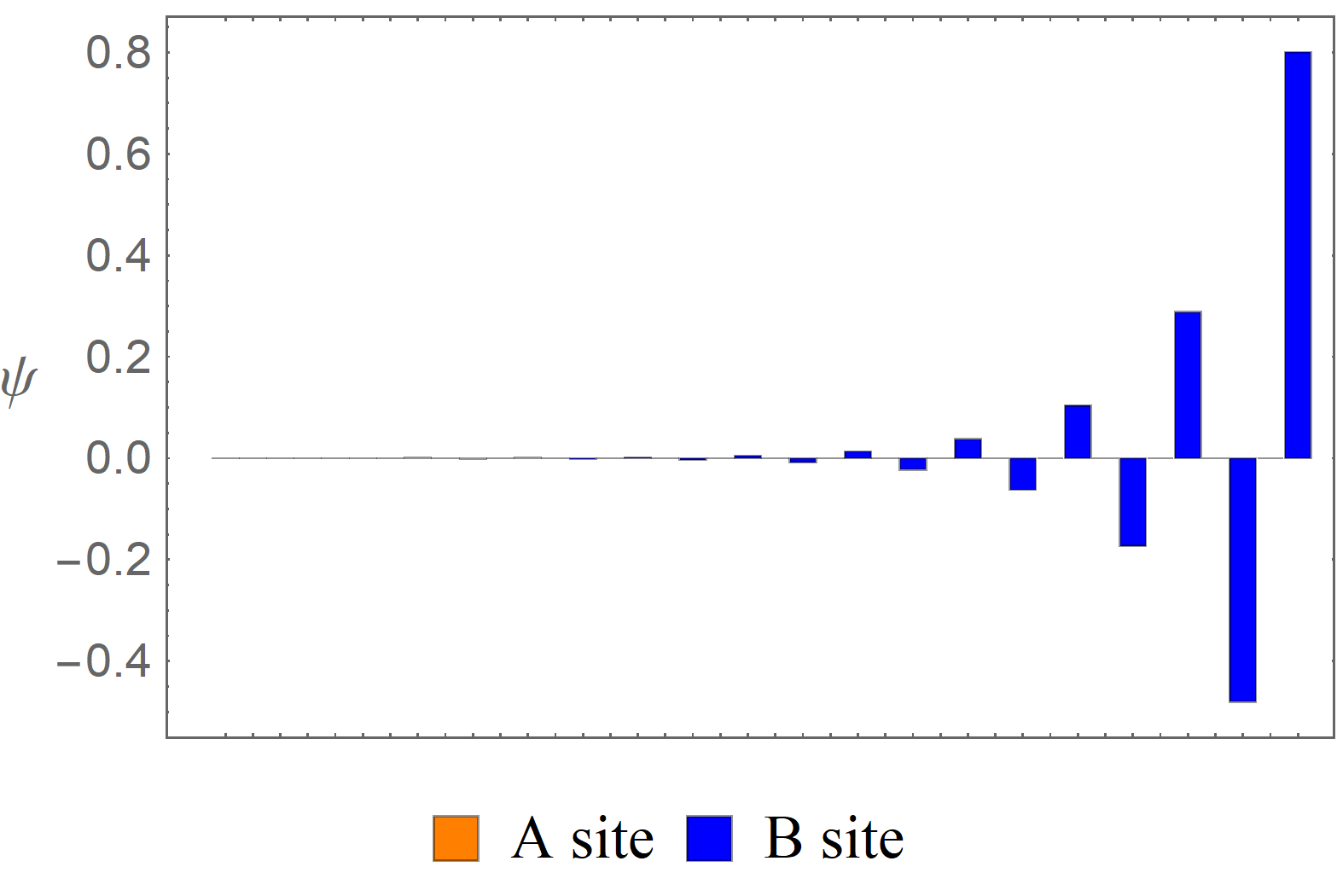}}
\caption{(a)The energy spectrum in the topological phase of the Rice-Mele model, where $(t_0, t_1, m_0)=(3, 5, 1)$. The energies of the left and right edge states are $m_0$, and $-m_0$, respectively. (b) The energy spectrum in the trivial phase, where $(t_0, t_1, m_0)=(5, 3, 1)$. (c), (d) The wave functions of the two edge states in the system.}\label{fig Rice-Mele}
\end{figure}

For a finite chain, we may again solve the recurrence relation when the number of sites is even.  We would obtain results similar to the SSH model:
\bea
&\;& \hskip -2.1cm E=\pm\sqrt{m_0^2+t_0^2+t_1^2+t_0 t_1\left(s+s^{-1}\right) }; \cr
&\;& \hskip -2.1cm t_0\left(\frac{s^{N+1}-s^{-N-1}}{s-s^{-1}} \right) +t_1 \left(\frac{s^{N}-s^{-N}}{s-s^{-1}} \right)=0.
\eea
When the number of sites is odd ($2N+1$), the recursion relation may again be solved analytically.  The bulk states are given by
\bea
&\;& \hskip -6.1cm s=e^{i\pi k/N},
\eea
with $k=1,\ldots, N,$ and
\bea
&\;& \hskip -2.1cm E_k =\pm\sqrt{m_0^2+t_0^2+t_1^2+2t_0 t_1\cos\left[\pi k/(N+1)\right] }.
\eea
For the edge state, we have $E=m_0$ and $s=-t_0/t_1$ analogous to the SSH model. We may also check the result numerically. Here, we will only consider the case that the number of sites to be 40 (even), since most of the results are quite similar to those of the SSH model. The energy spectrum for the case that $t_1>t_0$ and $t_1<t_0$ are shown in Figs.~\ref{fig Rice-Mele} (a) and (b). Indeed, edge states show up only in the case that $t_1>t_0$. The wave functions of the left and right edge states are shown in Figs.~\ref{fig Rice-Mele} (c) and (d), respectively and their energies are $m_0$, and $-m_0$.  This is consistent with the results obtained in Ref. \cite{Mong}.  Note that since the two edge states have different energies, their mixing here is negligible in contrast to the case of the SSH model.

So far, it seems all reminiscent of the SSH model. However, a 1D system without chiral symmetry is a trivial one according to the classification in the periodic table \cite{Periodic table}. Hence, there is no guarantee that the edge states are stable against perturbation. In fact, it has been shown in the literature that the Zak phase is not quantized in the Rice-Mele model where the chiral symmetry is violated \cite{Zak-phase_quantization}. On the other hand, we may relate the Rice-Mele model to a Chern insulator by introducing periodic time-dependence in the parameters \cite{Rice-Mele, Thouless}:
\bea
&\;& \hskip -2.5cm t_0(t) = \bar{t}_0+ \cos\left(2\pi t/T \right), \cr
&\;& \hskip -2.5cm t_1(t) =1, \cr
&\;& \hskip -2.5cm m_0(t) = \sin\left(2\pi t/T \right),
\eea
for example. As a matter of fact, the model is used to describe the charge-pumping process associated with the SSH model. If we consider $2\pi t/T$ as the momentum along an additional direction, then we see the Bloch Hamiltonian in Eq.~(\ref{RM}) would be equivalent to that of a Chern insulator, the Qi-Wu-Zhang (QWZ) model \cite{Rice-Mele}. In particular, by expanding it in terms of the Pauli matrices, i.e $\HH_{\rm QWZ} = h_i \t_i$, we have
\bea
&\;& \hskip -2.5cm h_1 = \left(\bar{t}_0 +\cos p_2 \right) + \cos p_1, \cr
&\;& \hskip -2.5cm h_2 = \sin p_1, \cr
&\;& \hskip -2.5cm h_3 = \sin p_2.  \label{Chern Insulator}
\eea
Following the convention we adupted, we define $h({\bf p}) = h_1({\bf p})+ i h_2({\bf p})$. It is well-known that we may determine the phase of the system by using the Chern number
\bea
\label{Chern number}
&\;& \hskip -2.5cm C = \frac{1}{2\pi} \int_{\rm BZ} d^2 p\;  B, \\
&\;& \hskip -2.5cm B = \pd_1 A_2 - \pd_2 A_1. \nn
\eea
Here, $B$ is the ``magnetic field'' and $A_a =\la \left. {\bf p}, - \right\vert  i \pd_a \left\vert {\bf p}, - \right.\ra$, the ``vector potential'', with $a=1,2$. The minus sign indicates that the energy eigenstates are in the lower band.  The wave functions of these eigenstates are known to be plagued by singularity at the north or south poles of the Bloch sphere.  As a result, we need two gauge patches to cover the whole sphere \cite{TI-model}:
\bea
\label{EgN}
&\;& \hskip -2.5cm \left|{\bf p},- \right\ra_{\rm I} = \frac{1}{{\cal N}_{\rm I}}\left(
\begin{matrix}
-\om + h_3  \cr
h \cr
\end{matrix}
\right), \;
\left|{\bf p},- \right\ra_{\rm II} = \frac{1}{{\cal N}_{\rm II}}\left(
\begin{matrix}
h^* \cr
-\om - h_3  \cr
\end{matrix}
\right).
\eea
Here,
\bea\label{NAB}
{\cal N}_{\rm I} =\sqrt{2\om(\om - h_3)}, \; {\cal N}_{\rm II} =\sqrt{2\om(\om + h_3)},
\eea
with $\om= \sqrt{h_1^2+ h_2^2+h_3^2}$. In gauge ${\rm I}$, the wave functions $\left|p,- \right\ra_{\rm I}$ is not well-defined at the north pole where $h =0$ and $h_3$ is positive; while in gauge ${\rm II}$, the wave function $\left|p,- \right\ra_{\rm II}$ is not well-defined at the south pole where $h =0$ and $h_3$ is negative.  These two wave functions are related by a phase factor: $\left|{\bf p},- \right\ra_{\rm II}  = e^{-i \phi({\bf p})} \left|{\bf p},- \right\ra_{\rm I}$, where, $\phi({\bf p}) =\arg(h).$  When the system is in the topological phase, both of the two gauge patches would be involved when we integrate over the whole Brillouin zone.  As a result, the Chern number defined in Eq.~(\ref{Chern number}) may be non-vanishing, even though a torus is free of boundaries. In particular, if we make use of the Stokes theorem, we would see that it receives contribution from the boundaries between the two gauge patches \cite{TI-model}. Note that since $h_3 = \sin p_2$, the north pole may only appear in the region that $0 \le p_2 \le \pi$ and the wave function $\left|{\bf p},- \right\ra_{\rm II}$ is always well-behaved in this region. On the other hand, the wave function $\left|{\bf p},- \right\ra_{\rm I}$ is always well-behaved in the region $-\pi \le p_2 \le 0$. This fact would be very useful for our derivation as we shall see later on.

By rotating along the axis $(\hat{x}_1+\hat{x}_2+\hat{x}_3)/\sqrt{3}$ by $2\pi/3$, we have
\bea
\label{QWZ}
\tilde{\HH}_{\rm QWZ}= \UU \HH_{\rm QWZ} \UU^\dag = \tilde{h}_i \t_i = \left(
\begin{matrix}
\tilde{h}_3({\bf p}) & \tilde{h}^*({\bf p}) \cr
\tilde{h}({\bf p}) & -\tilde{h}_3({\bf p}) \cr
\end{matrix}
\right).
\eea
Here, $\tilde{h}_1= h_2$, $\tilde{h}_2= h_3$, $\tilde{h}=h_2+i h_3$, $\tilde{h}_3= h_1$, and
$\UU= \frac{1}{2} \left(
\begin{matrix}
1+i & 1+i \cr
-1+i & 1-i \cr
\end{matrix}
\right).$
In terms of $\tilde{\HH}_{\rm QWZ}$, we have the following energy eigenstates in the lower band
\bea
\label{EgNt}
&\;& \hskip -2.5cm \left|{\bf p},- \right\rra_{\rm \,I} = \frac{1}{\tilde{\cal N}_{\rm I}}\left(
\begin{matrix}
-\om + \tilde{h}_3  \cr
\tilde{h} \cr
\end{matrix}
\right), \;
\left|{\bf p},- \right\rra_{\rm \,II} = \frac{1}{\tilde{\cal N}_{\rm II}}\left(
\begin{matrix}
\tilde{h}^* \cr
-\om - \tilde{h}_3  \cr
\end{matrix}
\right),
\eea
with
\bea\label{NABt}
\tilde{\cal N}_{\rm I} =\sqrt{2\om(\om - \tilde{h}_3)}, \; \tilde{\cal N}_{\rm II} =\sqrt{2\om(\om + \tilde{h}_3)}.
\eea
Since $\tilde{h}_1^2+\tilde{h}_2^2+\tilde{h}_3^2 = h_1^2+h_2^2+h_3^2$, $\om$ remains the same.  Again, these two wave functions of the eigenstates are singular at the new north and south poles, where $\tilde{h} =0$ while $\tilde{h}_3$ is positive and negative, respectively. Similarly, the two wave functions are related by a phase factor: $\left|{\bf p},- \right\rra_{\rm \,II}  = e^{-i \tilde{\phi}({\bf p})} \left|{\bf p},- \right\rra_{\rm \,I}$, with $\tilde{\phi} ({\bf p})=\arg(\tilde{h}).$

Some comments are in order. First, it is known that $B=\epsilon_{ijk} h_i \frac{\pd h_j}{\pd p_1} \frac{\pd h_k}{\pd p_2}/\left(2 \om^3\right)$. Thus, it makes no difference whether we use $\HH_{\rm QWZ}$ or $\tilde{\HH}_{\rm QWZ}$ to calculate the magnetic field and Chern number. Second, since $\left|{\bf p},-\, \right\rra$ and $\UU \left|{\bf p},- \right\ra$ are both energy eigenstates of $\tilde{\HH}_{\rm QWZ}$ with the same energy, $-\om$, they are also related by a phase factor $\left|{\bf p},-\, \right\rra = \e^{i\theta({\bf p})} \UU \left|{\bf p},- \right\ra$.  Taking advantage of these facts, we may hereafter choose whichever Hamiltonian and eigenstate that are more convenient for us to carry out the calculation or argument.

By using $\tilde{\HH}_{\rm QWZ}$,  it is easy to see that $\tilde{h}$ is vanishing at the four TRIM points ${\bf P}_1= (0,0), {\bf P}_2=(\pi, 0), {\bf P}_3=(0, \pi)$ and ${\bf P}_4=(\pi, \pi)$.  For the Chern insulator that we are considering, if the signs of $\tilde{h}_3$ at these four points are all the same then the wave function may be defined globally and the system would be in the trivial phase. In contrast, if one of the signs of  $\tilde{h}_3({\bf P}_i)$ is different from others, then two gauge patches are needed to define the wave function and the systems would be in the topological phase.  In other words, one may find out the phase of the system by calculating $\Pi_{i=1}^{4}  \tilde{h}_3({\bf P}_i)$. If the sign of the product is negative, then it is in the topological phase; otherwise it is in the trivial phase.  Carrying out the calculation explicitly, one sees that this leads to the condition
\be
\label{topological CI}
-2< \bar{t}_0<2.
\ee
In contrast, if $|\bar{t}_0|>2$ then the system is in the trivial phase.

In light of the relation between the SSH model and Rice-Mele model which is in turn related to a Chern insulator, we would like to find the connection between the 2D Chern number of a QWZ model to the Zak phase of the associated SSH model by considering the following momentum strip in the Brillouin zone:
\be
p_1 \in [0, 2\pi), 0 \le p_2 \le \pi.
\ee
Since the Berry curvature is even in $p_2$, if we integrate it over the strip the result would be $C/2$. Making use of the Stokes theorem, we see that
\be
\label{Chern and Zak}
\g( 0) - \g( \pi) = \pi C.
\ee
Here, the Zak phase for a specific $p_2$ is given by
\bea
&\;& \hskip -2.1cm \g(p_2) = \int_0^{2\pi} dp_1 \la \left. {\bf p}, - \right\vert  i \pd_1 \left\vert {\bf p}, - \right.\ra,
\eea
and it is closely related to the Berry phase \cite{Berry}.  Here, we use the wave function in Eq.~(\ref{EgN}), since the corresponding Zak phase is exactly in the form that it is defined in the literature \cite{Zak}. Note that as $h_3 = \tilde{h}_2 = 0$ at $p_2 = 0, \pi$, chiral symmetry is restored along these axes and thus the Zak phase is quantized, which is proportional to the winding number: $\g(0) = \pi \n(0)$, and $ \g(\pi) = \pi \n(\pi)$.  Moreover, since $h_3 = \sin p_2$, which is always positive in the momentum strip that we are considering, we may use the wave function $\left|{\bf p},- \right\ra_{\rm II}$ to do the calculation safely. Combining these results with Eq.~(\ref{Chern and Zak}), we achieve
\be
\label{Chern and Winding number}
\n(0)-\n(\pi)  =C.
\ee
Therefore, the 2D Chern number of the system is connected to the difference in the 1D winding number of the SSH models associated with the Chern insulator.  An expression similar to the above one has been mentioned in Ref. \cite{Z2} for the $Z_2$ case.

If we consider a generic momentum strip in the Brillouin zone instead :
\be
p_1 \in [0, 2\pi), 0 \le p_2 \le \tilde{p} \le \pi,
\ee
then we have
\be
\g(0)-\g(\tilde{p})  = \frac{1}{2\pi} \int_{0}^{\tilde{p}} d p_2\; \int_{0}^{2\pi} d p_1\;  B.
\ee
Since the chiral symmetry is lost except at $p_2 =0, \pi$, the above identity provides another perspective to understand why $\g(\tilde{p})$ is generally not quantized.

Let's use the Chern insulator that we mentioned above as an example to verify the identity in Eq.~({\ref{Chern and Winding number}). It can be seen easily that $C$ is non-vanishing if and only if $\n(0)$ and $\n( \pi)$ have different values.  As $h_1(p_1,0)=  \left(\bar{t}_0 +1 \right) + \cos p_1$ and $h_1(p_1,\pi)=  \left(\bar{t}_0 -1 \right) + \cos p_1$,  we see that the Chern insulator is in the topological phase when
\be
\label{Different Zak phase}
|\bar{t}_0 +1|<1, |\bar{t}_0 -1|>1\; {\rm or}\;  |\bar{t}_0 +1|>1, |\bar{t}_0 -1|<1.
\ee
After some algebra, it may be shown that the above condition is equivalent to the one that the QWZ model is in the topological phase, given in Eq.~(\ref{topological CI}).

As long as the Berry curvature remains to be an even function of $p_2$, similar generalization may also be made in the case of the extended SSH models.  It has been shown in Eq.~(\ref{ssh ext1 hamiltonian tilde}) that type 1 and type 2 extended SSH models are in fact equivalent if the periodic boundary condition is imposed. Thus, we will only consider the generalization of type 1 extended SSH model. To be concrete, let's be more specific and consider the case that $t_1$ and $t_2$ are independent of $p_2$: $t_0(p_2) = \bar{t}_0 +\cos p_2$, $t_1=\bar{t}_1$ and $t_2= \bar{t}_2$ so that $ t_0(0)=  \bar{t}_0 +1$ and $t_0(\pi)=  \bar{t}_0 -1$.  Consequently, the Bloch Hamiltonian of the extended QWZ model is given by
\bea
&\;& \hskip -2.5cm h_1 = \left(\bar{t}_0 +\cos p_2 \right) + \bar{t}_1\cos p_1 + \bar{t}_2 \cos (2p_1), \cr
&\;& \hskip -2.5cm h_2 = \bar{t}_1\sin p_1+ \bar{t}_2 \sin (2p_1), \cr
&\;& \hskip -2.5cm h_3 = \sin p_2.  \label{Extended Chern Insulator}
\eea
For convenience, we choose $\bar{t}_1$ to be positive without loss of generality. From the results of the extended SSH models, we have
\bea
\label{Winding 0}
&\;& \hskip -2.5cm \n(0)=
\begin{cases}
2, & \mbox{for } |\bar{t}_2 + \bar{t}_0 +1| > \bar{t}_1, \mbox{ and }  \bar{t}_2 - |\bar{t}_0+1| >0 \\
1, & \mbox{for } |\bar{t}_2 + \bar{t}_0 +1| < \bar{t}_1; \\
0, & \mbox{for } |\bar{t}_2 + \bar{t}_0 +1| > \bar{t}_1, \mbox{ and }  \bar{t}_2 - |\bar{t}_0+1| <0.
\end{cases}  \\
\label{Winding Pi}
&\;& \hskip -2.5cm \n(\pi)=
\begin{cases}
2, & \mbox{for } |\bar{t}_2 + \bar{t}_0-1| > \bar{t}_1, \mbox{ and }  \bar{t}_2 - |\bar{t}_0-1| >0; \\
1, & \mbox{for } |\bar{t}_2 + \bar{t}_0-1| < \bar{t}_1; \\
0, & \mbox{for } |\bar{t}_2 + \bar{t}_0-1| > \bar{t}_1, \mbox{ and }  \bar{t}_2 - |\bar{t}_0+1| <0.
\end{cases}
\eea
Therefore, we can obtain a Chern insulator with its Chern number $C$ ranging from -2 to 2 by choosing suitable values of $\bar{t}_0$ and $\bar{t}_2$.

To verify the identity in Eq.~({\ref{Chern and Winding number}) for the system, we must also calculate the Chern number. Again, it is easier to use $\tilde{\HH}_{\rm QWZ}$ for such a task. We will first find all the zeroes of $\tilde{h}({\bf p})$ and then determine the signs of $\tilde{h}_3({\bf p})$ and vorticities of $\tilde{h}({\bf p})$ around these points. Notice that $\tilde{h}_1=h_2=\sin p_1 \left( \bar{t}_1+2 \bar{t}_2 \cos p_1 \right)$ and $\tilde{h}_2=h_3=\sin p_2$. Thus, it is obvious that $\tilde{h}({\bf p})$ is again vanishing at the TRIM points. After some straight forward calculation, we obtain
\bea
\hskip -2.5cm
\begin{cases}
\tilde{h}_3({\bf P}_1) = \bar{t}_0 +\bar{t}_2 + \bar{t}_1+1, \\
\tilde{h}_3({\bf P}_2) = \bar{t}_0 +\bar{t}_2 - \bar{t}_1+1, \\
\tilde{h}_3({\bf P}_3) = \bar{t}_0 +\bar{t}_2 + \bar{t}_1-1, \\
\tilde{h}_3({\bf P}_4) = \bar{t}_0 +\bar{t}_2 - \bar{t}_1-1,
\end{cases}  \quad \mbox{and} \quad
\begin{cases}
\tilde{h}({\bf p}) \approx (2\bar{t}_2+\bar{t}_1)\Delta p_1 + i \Delta p_2, \\
\tilde{h}({\bf p}) \approx (2\bar{t}_2-\bar{t}_1)\Delta p_1 + i \Delta p_2, \\
\tilde{h}({\bf p}) \approx (2\bar{t}_2+\bar{t}_1)\Delta p_1 - i \Delta p_2, \\
\tilde{h}({\bf p}) \approx (2\bar{t}_2-\bar{t}_1)\Delta p_1 - i  \Delta p_2.
\end{cases}
\eea
When $\bar{t}_2 > \bar{t}_1/2$, $\tilde{h}({\bf p})$ would have more zeroes at ${\bf P}_5 =(p_0,0)$, ${\bf P}_6 =(-p_0,0)$, ${\bf P}_7 =(p_0,\pi)$, and ${\bf P}_8 =(-p_0,\pi)$. Here, $p_0 = \cos^{-1}\left[-\bar{t}_1/(2\bar{t}_2) \right]$. Around these additional zeroes, we have
\bea
\hskip -2.5cm
\begin{cases}
\tilde{h}_3({\bf P}_5) = \bar{t}_0 -\bar{t}_2 +1, \\
\tilde{h}_3({\bf P}_6) = \bar{t}_0 -\bar{t}_2+1, \\
\tilde{h}_3({\bf P}_7) = \bar{t}_0 -\bar{t}_2 -1, \\
\tilde{h}_3({\bf P}_8) = \bar{t}_0 -\bar{t}_2 -1,
\end{cases}  \quad \mbox{and} \quad
\begin{cases}
\tilde{h}({\bf p}) \approx \left[\bar{t}_1^2/(2\bar{t}_2) - 2\bar{t}_2 \right] \Delta p_1 + i \Delta p_2, \\
\tilde{h}({\bf p}) \approx \left[\bar{t}_1^2/(2\bar{t}_2) - 2\bar{t}_2 \right] \Delta p_1 + i \Delta p_2, \\
\tilde{h}({\bf p}) \approx \left[\bar{t}_1^2/(2\bar{t}_2) - 2\bar{t}_2 \right] \Delta p_1 - i \Delta p_2, \\
\tilde{h}({\bf p}) \approx \left[\bar{t}_1^2/(2\bar{t}_2) - 2\bar{t}_2 \right] \Delta p_1 - i \Delta p_2.
\end{cases}
\eea
To find the Chern number of the system, we must divide the parameters space into the following three categories.
\bde
\item{1.)} $|\bar{t}_2|<\bar{t}_1/2$:

We only need to take into account of the contributions from the TRIM points.  If $|\bar{t}_0 + \bar{t}_2 |>\bar{t}_1+1$, all the $\tilde{h}_3$ at the TRIM points are of the same sign.
\bde
\item{i.)}$\bar{t}_1>1$:

If $\bar{t}_1-1<\bar{t}_0 + \bar{t}_2<\bar{t}_1+1$, then only $\tilde{h}_3({\bf P}_4)$ is negative; if $-(\bar{t}_1+1)<\bar{t}_0 + \bar{t}_2<-(\bar{t}_1-1)$, then only $\tilde{h}_3({\bf P}_1)$ is positive. Meanwhile, if $-(\bar{t}_1-1)< \bar{t}_0 + \bar{t}_2<\bar{t}_1-1$, then only $\tilde{h}_3({\bf P}_4)$ and $\tilde{h}_3({\bf P}_2)$ are negative.

\item{ii.)}$0<\bar{t}_1<1$:

If $1-\bar{t}_1<\bar{t}_0+ \bar{t}_2<\bar{t}_1+1$, then only $\tilde{h}_3({\bf P}_4)$ is negative; if $-(\bar{t}_1+1)<\bar{t}_0 + \bar{t}_2<-(1-\bar{t}_1)$, then only $\tilde{h}_3({\bf P}_1)$ is positive. Meanwhile, if $-(1-\bar{t}_1)< \bar{t}_0 + \bar{t}_2<1-\bar{t}_1$, then only $\tilde{h}_3({\bf P}_4)$ and $\tilde{h}_3({\bf P}_3)$ are negative.
\ede
Since the vorticities of $\tilde{h}({\bf p})$ are positive around ${\bf P}_1$ and ${\bf P}_4$ but negative around ${\bf P}_2$ and ${\bf P}_3$ under the condition $|\bar{t}_2|<\bar{t}_1/2$, we have
\bea
\label{Chern I}
&\;& \hskip -2.5cm C =
\begin{cases}
\;\;1, & \mbox{for } -(\bar{t}_1+1)<\bar{t}_0 + \bar{t}_2 <-|\bar{t}_1-1|; \\
\;\;0, & \mbox{for } |\bar{t}_0 + \bar{t}_2| >\bar{t}_1+1 \mbox{ or } |\bar{t}_0 + \bar{t}_2| <|\bar{t}_1-1|; \\
-1, & \mbox{for } |\bar{t}_1-1|<\bar{t}_0 + \bar{t}_2 <\bar{t}_1+1.
\end{cases}
\eea
\ede

\bde
\item{2.)} $\bar{t}_2>\bar{t}_1/2$:

We need to take into account of the contributions from all the eight points.  By analyzing the signs of $\tilde{h}_3({\bf p})$ and vorticities of $\tilde{h}({\bf p})$ around ${\bf P}_1, \ldots, {\bf P}_8$, we arrive at
\bde
\item{i.)}$\bar{t}_1\ge1$:
\bea
\label{Chern II}
&\;& \hskip -2.5cm C =
\begin{cases}
\;\;1, & \mbox{for } \bar{t}_1-1<|\bar{t}_0 + \bar{t}_2 |<\bar{t}_1+1, \bar{t}_0 - \bar{t}_2<-1; \\
\;\;0, & \mbox{for } |\bar{t}_0 + \bar{t}_2| >\bar{t}_1+1, |\bar{t}_0 - \bar{t}_2|>1 \mbox{ or } |\bar{t}_0 + \bar{t}_2| <\bar{t}_1-1,   \bar{t}_0 - \bar{t}_2<-1; \\
-1, & \mbox{for } \bar{t}_1-1<\bar{t}_0 + \bar{t}_2 <\bar{t}_1+1, |\bar{t}_0- \bar{t}_2| <1; \\
-2, & \mbox{for } \bar{t}_0 + \bar{t}_2 >\bar{t}_1+1, |\bar{t}_0- \bar{t}_2| <1.
\end{cases}
\eea

\item{ii.)}$0\le\bar{t}_1<1$:
\bea
\label{Chern III}
&\;& \hskip -2.5cm C =
\begin{cases}
\;\;2, & \mbox{for } |\bar{t}_0 + \bar{t}_2| <1-\bar{t}_1, \bar{t}_0 - \bar{t}_2<-1; \\
\;\;1, & \mbox{for } 1-\bar{t}_1<|\bar{t}_0 + \bar{t}_2| <\bar{t}_1+1, \bar{t}_0 - \bar{t}_2<-1; \\
\;\;0, & \mbox{for } |\bar{t}_0 + \bar{t}_2| >\bar{t}_1+1, |\bar{t}_0 - \bar{t}_2|>1 \mbox{ or } |\bar{t}_0 + \bar{t}_2| <1-\bar{t}_1,  |\bar{t}_0 - \bar{t}_2|<1; \\
-1, & \mbox{for } 1-\bar{t}_1<\bar{t}_0 + \bar{t}_2 <\bar{t}_1+1, |\bar{t}_0- \bar{t}_2| <1; \\
-2, & \mbox{for } \bar{t}_0 + \bar{t}_2 >\bar{t}_1+1, |\bar{t}_0- \bar{t}_2| <1.
\end{cases}
\eea
\ede
\ede

\bde
\item{3.)} $\bar{t}_2<-\bar{t}_1/2$:

Again, we need to take into account of the contributions from all the eight points.  By similar analysis, we arrive at
\bde
\item{i.)}$\bar{t}_1\ge1$:
\bea
\label{Chern IV}
&\;& \hskip -2.5cm C =
\begin{cases}
\;\;2, & \mbox{for } \bar{t}_0 + \bar{t}_2 <-(\bar{t}_1+1), |\bar{t}_0- \bar{t}_2| <1; \\
\;\;1, & \mbox{for } -(\bar{t}_1+1)<\bar{t}_0 + \bar{t}_2 <-(\bar{t}_1-1), |\bar{t}_0- \bar{t}_2| <1; \\
\;\;0, & \mbox{for } |\bar{t}_0 + \bar{t}_2| >\bar{t}_1+1, |\bar{t}_0 - \bar{t}_2|>1 \mbox{ or } |\bar{t}_0 + \bar{t}_2| <\bar{t}_1-1,  \bar{t}_0 - \bar{t}_2>1; \\
-1, & \mbox{for } \bar{t}_1-1<|\bar{t}_0 + \bar{t}_2 |<\bar{t}_1+1, \bar{t}_0 - \bar{t}_2>1.
\end{cases}
\eea

\item{ii.)}$0\le\bar{t}_1<1$:
\bea
\label{Chern V}
&\;& \hskip -2.5cm C =
\begin{cases}
\;\;2, & \mbox{for } \bar{t}_0 + \bar{t}_2 <-(\bar{t}_1+1), |\bar{t}_0- \bar{t}_2| <1; \\
\;\;1, & \mbox{for } -(\bar{t}_1+1)<\bar{t}_0 + \bar{t}_2 <-(\bar{t}_1-1), |\bar{t}_0- \bar{t}_2| <1; \\
\;\;0, & \mbox{for } |\bar{t}_0 + \bar{t}_2| >\bar{t}_1+1, |\bar{t}_0 - \bar{t}_2|>1 \mbox{ or } |\bar{t}_0 + \bar{t}_2| <1-\bar{t}_1,  |\bar{t}_0 - \bar{t}_2|<1; \\
-1, & \mbox{for } 1-\bar{t}_1<|\bar{t}_0 + \bar{t}_2| <\bar{t}_1+1, \bar{t}_0 - \bar{t}_2>1; \\
-2, & \mbox{for } |\bar{t}_0 + \bar{t}_2| <1-\bar{t}_1, \bar{t}_0 - \bar{t}_2>1.
\end{cases}
\eea
\ede
\ede
It is easy to check explicitly that the results summarized from Eq.~(\ref{Chern I}) to Eq.~(\ref{Chern V}) are consistent with those in Eqs.~(\ref{Winding 0}) and (\ref{Winding Pi}). Using the above results, we show the phase diagrams of the extended QWZ model for two typical cases with $\bar{t}_1 = 1.5, 0.5$ in Fig.~\ref{fig Ext-QWZ}.
\begin{figure}[h!]
\centering
\subfloat[]{\includegraphics[width=0.45\textwidth]{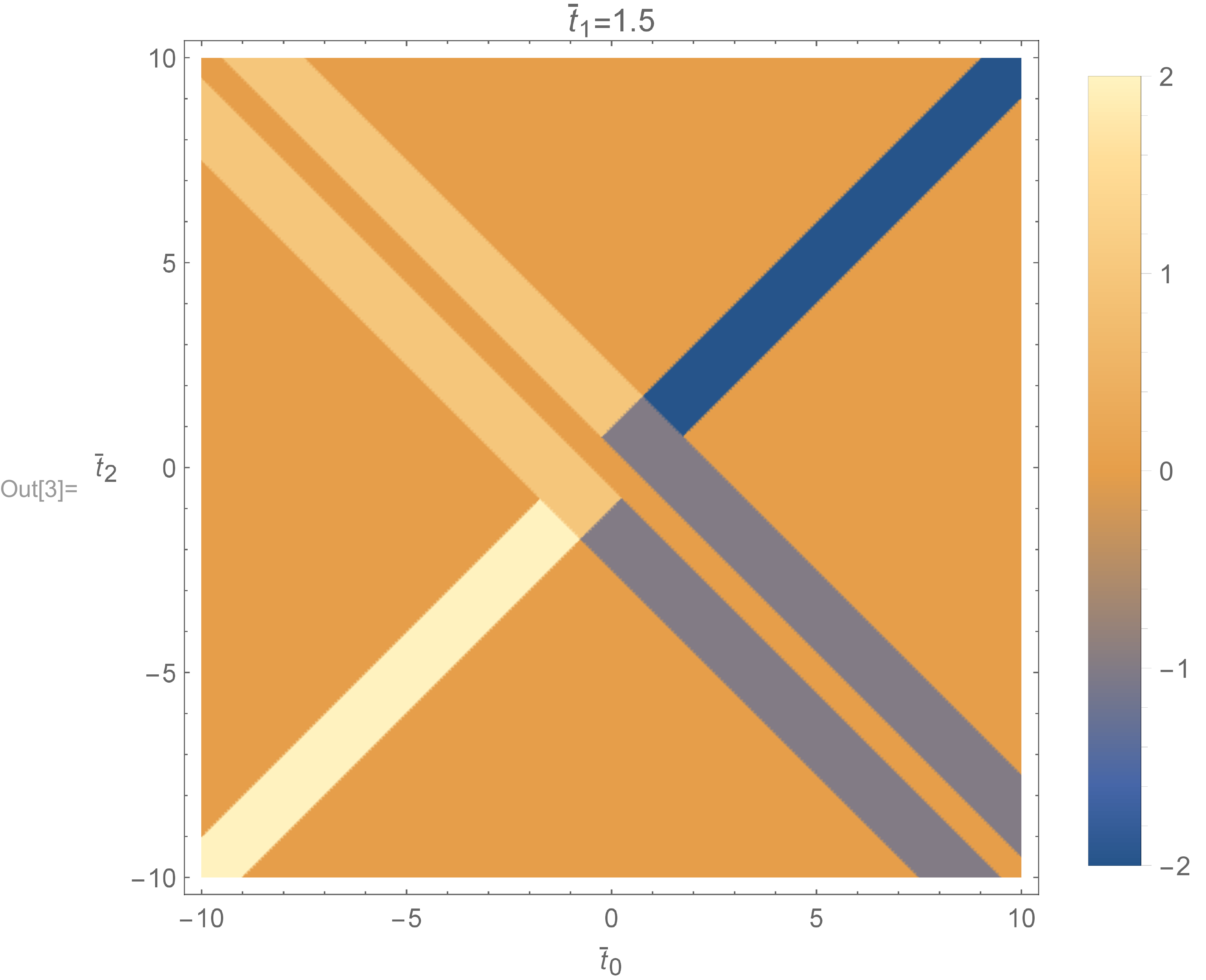}}
\subfloat[]{\includegraphics[width=0.45\textwidth]{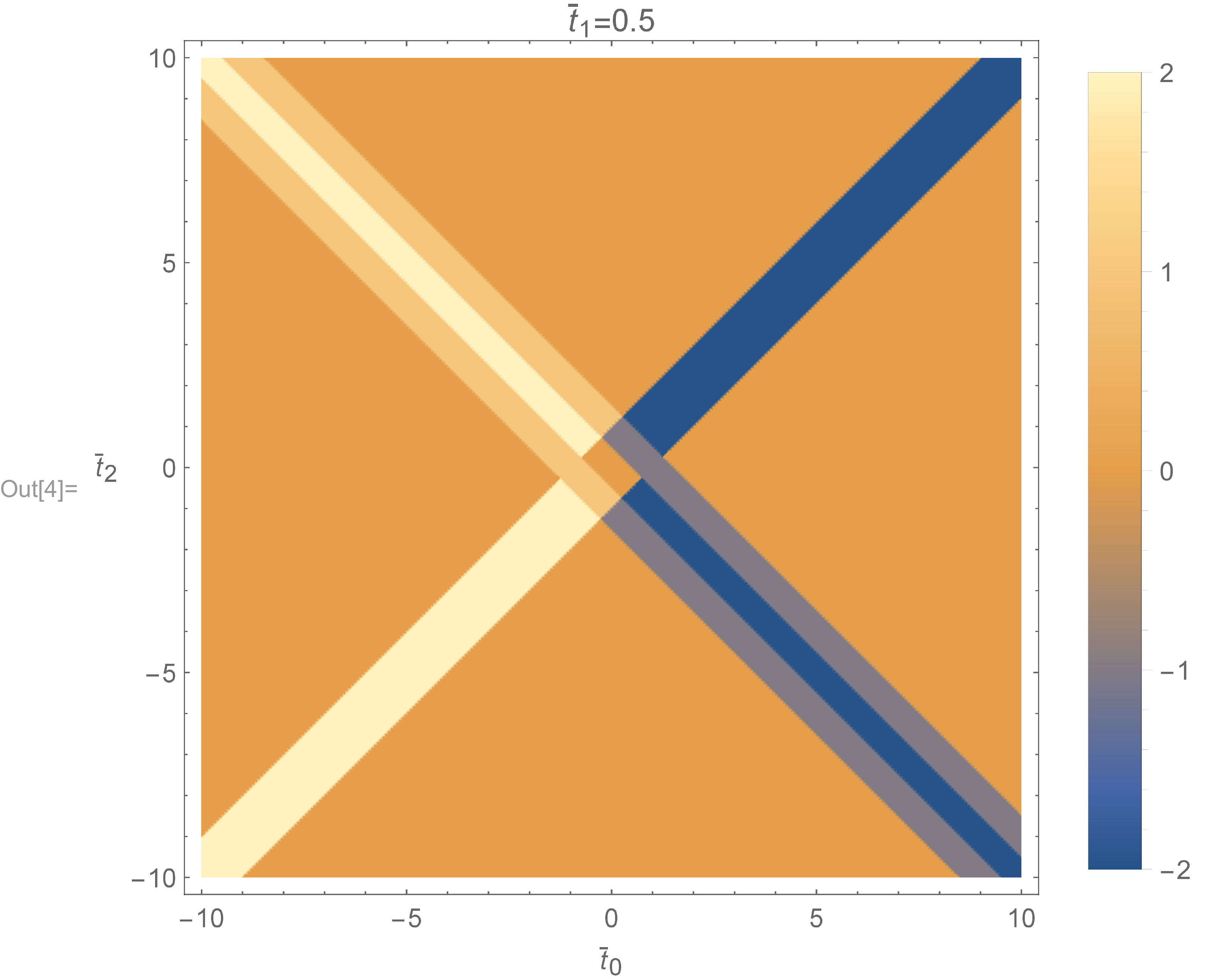}}
\caption{(a)The phase diagram for  $\bar{t}_1 = 1.5$. (b)The phase diagram for  $\bar{t}_1 =  0.5$.}\label{fig Ext-QWZ}
\end{figure}

Some comments are in order.  First, even if $\tilde{h}_3({\bf p})$ is not an even function of $p_2$, certain relation between the 1D winding number and 2D Chern number may still exist as long as there exist two different momenta $p_2 = p_{2{\rm A}}, p_{2{\rm B}}$ where that the chiral symmetry is restored, i.e. $\tilde{h}_3(p_1,  p_{2{\rm A}})=\tilde{h}_3(p_1,  p_{2{\rm B}})=0$. To be more specific, let's assume that $\tilde{h}_3({\bf p})$ is positive in the momentum region $p_{2{\rm A}} < p_2 < p_{2{\rm B}}$ and negative in the region $p_{2{\rm B}} < p_2 < 2\pi+p_{2{\rm B}}$. Then, the wave function in gauge patch ${\rm II}$ and ${\rm I}$ would be well-defined throughout the former region and the latter region, respectively. Thus, we may apply Stokes' theorem to the two regions separately and obtain
\bea
&\;& \hskip -2.5cm \frac{1}{2}\left\{\n_{\rm II}(p_{2 \rm A})- \n_{\rm II}(p_{2 \rm B})\right\}  =C_{\rm II}; \\
&\;& \hskip -2.5cm \frac{1}{2}\left\{\n_{\rm I}(p_{2 \rm B})- \n_{\rm I}(p_{2 \rm A})\right\}  =C_{\rm I}.
\eea
Since the Chern number of the system $C= C_{\rm I}+C_{\rm II}$, we may still relate it to the 1D winding numbers at $p_{2{\rm A}}$ and $p_{2{\rm A}}$. Following similar argument, we expect this may be generalized to more generic Chern insulators.  Second, by integrating $\nabla\cdot {\bf B}$ over a suitable region of the Brillouin zone in 3D, we may obtain a similar identity between the number of Weyl points and the difference in the 2D Chern numbers on the boundary surfaces. This is reminiscent of the classification of the 3D strong topological insulators \cite{Review, STI}.

\subsection{IV. Conclusion and discussion}

In this paper, we first show that if we choose properly the unit cells according to the left and right boundaries in a finite SSH chain then the bulk-edge correspondence always holds. In particular, the winding numbers $\n$ corresponding to the two unit cells may be used to predict the numbers of edge states on the left and right boundaries, respectively.  We then show that this may be generalized to the extended SSH models where there are next to nearest neighbor hopping amplitudes. We demonstrate that these results may be used to understand the edge states of CNT's. It is worth mentioning that the existence of edge states depend sensitively on the boundary conditions. Finally, using the idea of charge pumping we generalize the extended SSH models to the extended QWZ model with the Chern numbers ranging from $-2 $ to $2$.  We also establish a relation between Chern number and the difference in the 1D winding numbers by integrating the magnetic over the momentum region $0\le p_2 \le \pi, 0 \le p_1< 2\pi$ in the Brillouin zone. This relation may be used as a guiding principle to construct generic Chern insulators from 1D topological insulators.  A similar relation may also exist in higher dimensions.

We would like to discuss some possible directions for further study.  We show in Sec. II that the bulk edge correspondence may be used to understand when and how the edge states in a CNT with various edges would appear.  It is likely that similar understanding may be generalized to the counting of the number of edge states in a carbon nano-ribbon, where there are even more boundaries \cite{CNR}.

As we mentioned, there is a relation between the number of Weyl points and the difference in 2D Chern numbers on the boundary surfaces in 3D. This is reminiscent of the rules that we use to determine whether a 3D topological insulator is a strong one or not.  However, this seems to suggest that there always exist Weyl points in a strong topological insulator.  It would be interesting to carry out more detailed study along this line.

\section*{Acknowledgments}
The work is supported in part by the Grants 107-2112-M-003-009 of the Ministry of Science and Technology, Taiwan.  The authors would like to thank Prof. Ming-Che Chang for his illuminating lectures and stimulating discussions.

\appendix
\section{Appendix A: Numerical results of the type 1 extended Rice-Mele model}\label{ER-M}

For completeness, let's show some numerical results of a finite chain of the type 1 extended Rice-Mele model. Similar to the case of SSH model, most of the results in the extended SSH model may be carried over to the extended Rice-Mele model, except that the energy of edge states are now shifted to $E = m_0$ and $-m_0$, respectively.  Again, the edge states on the left and right boundaries have little mixing since their energies are different here. For convenience, we will refer to the extended Rice-Mele model by the winding number $\n$ of the associated extended SSH model from now on. We will consider the case that $\n=2$ and there are 40 (even) sites for illustration.  The energy spectrum and the associated wave functions of all the edge states are shown in Fig.~\ref{fig ext-RM even}.

Similar to the case for the type 1 extended SSH model, by varying the parameters $t_0, t_1,$ and $t_2$ of the system, we may achieve systems corresponding to other values of $\n$.  As most of the results are quite similar to those of the associated extended SSH model, we would not show them here.

\vfill\eject

\begin{figure}[t!]
\centering
\subfloat[]{\includegraphics[width=0.50\textwidth]{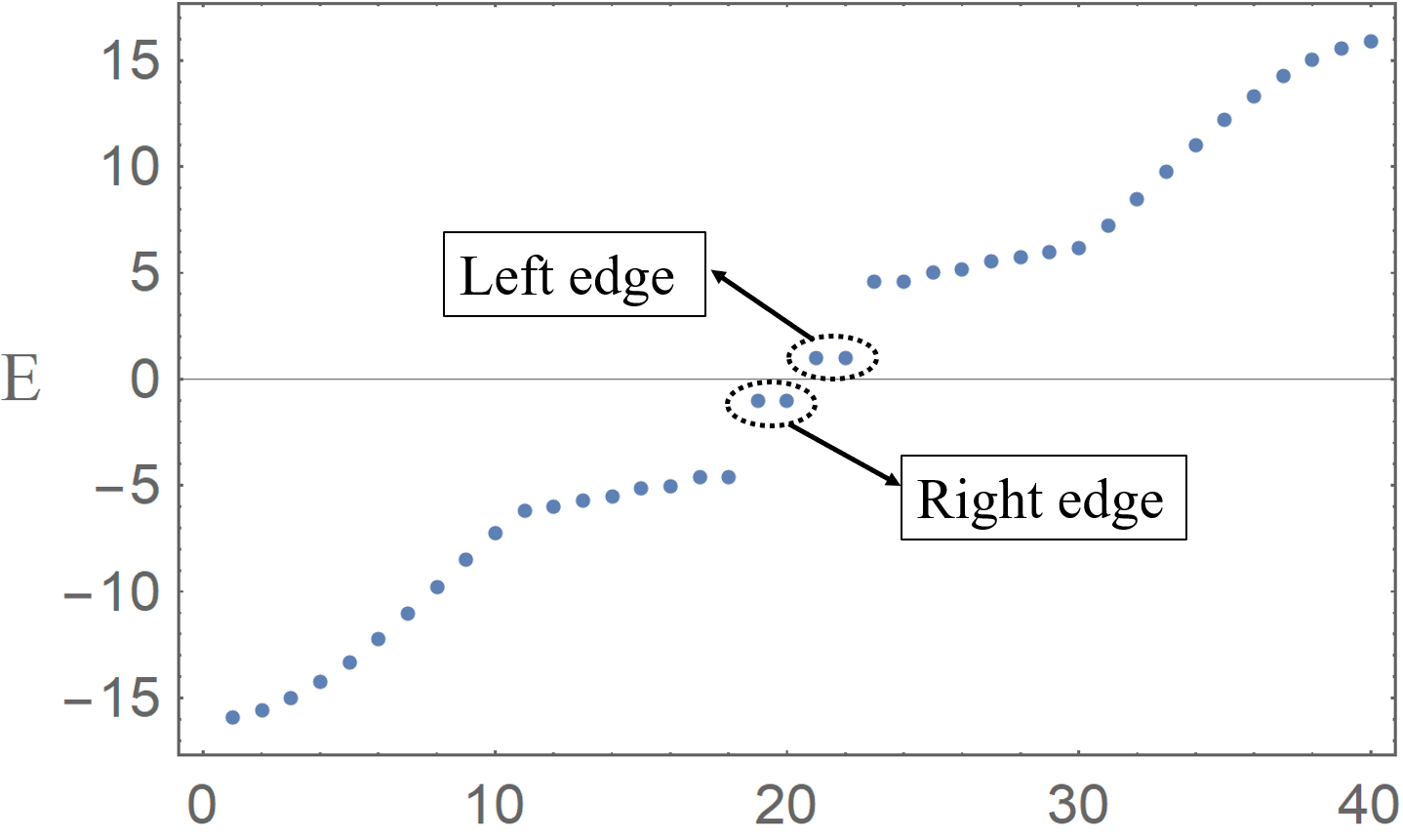}}\\
\subfloat[]{\includegraphics[width=0.30\textwidth]{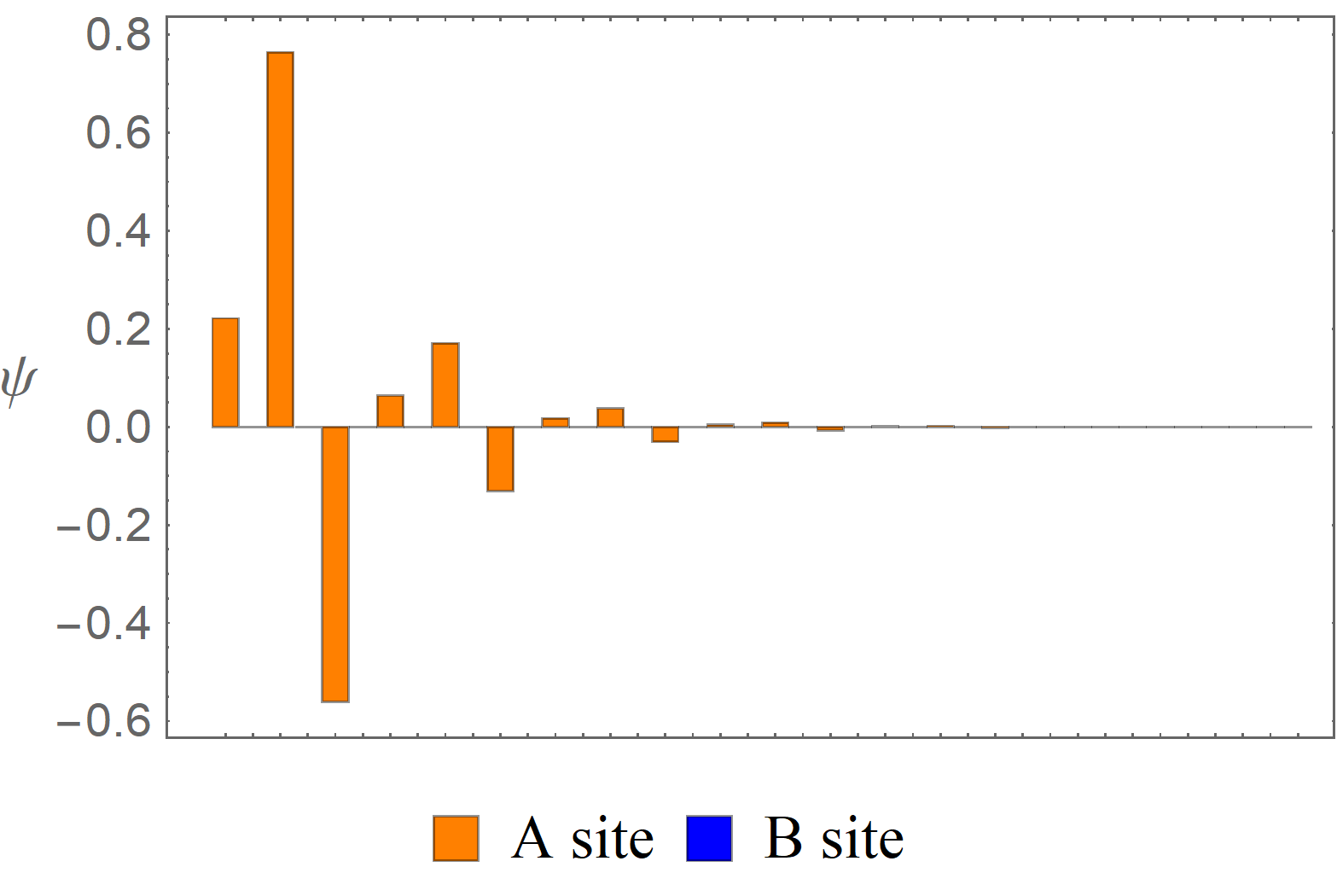}}\hskip 0.5cm
\subfloat[]{\includegraphics[width=0.30\textwidth]{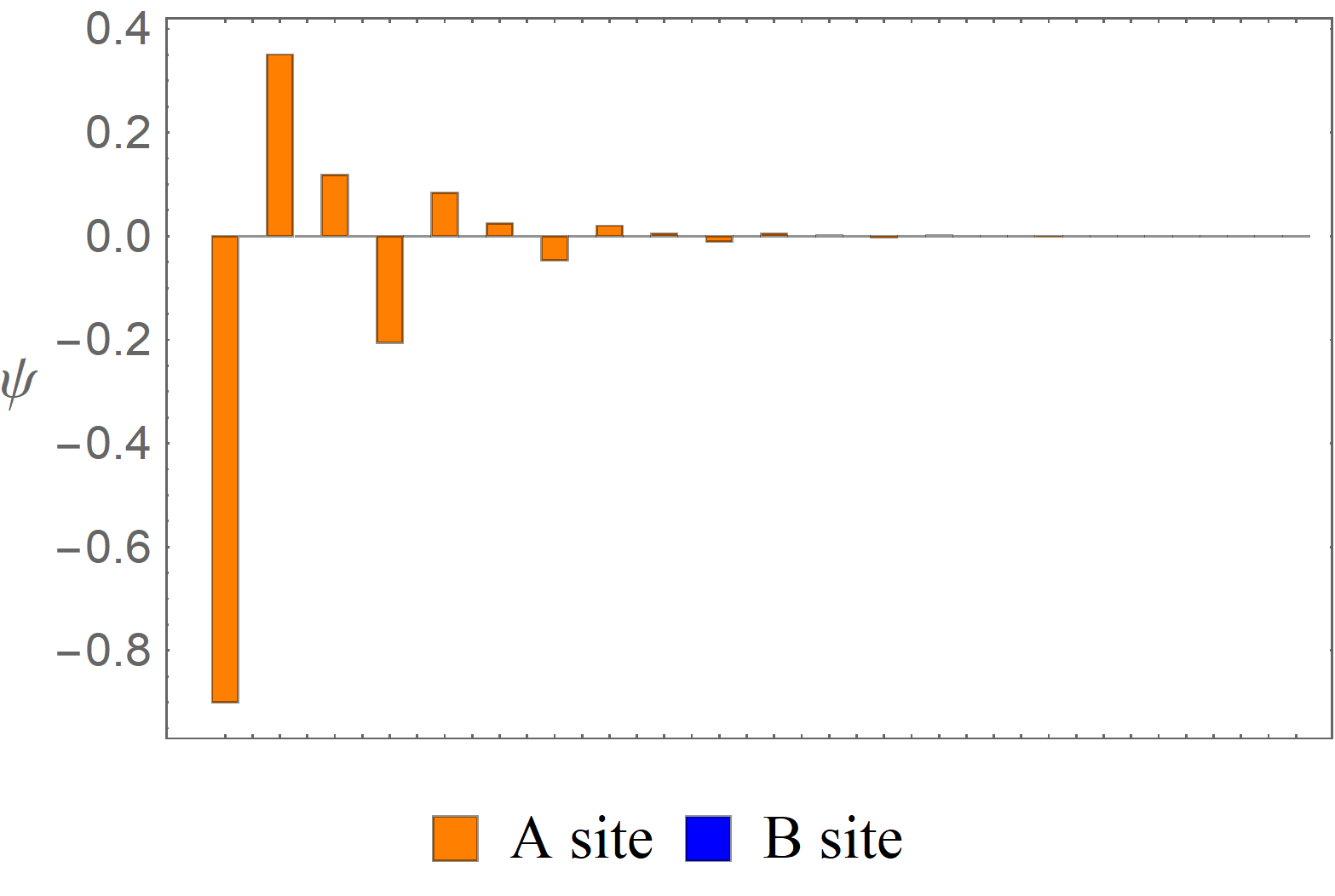}}\\
\subfloat[]{\includegraphics[width=0.30\textwidth]{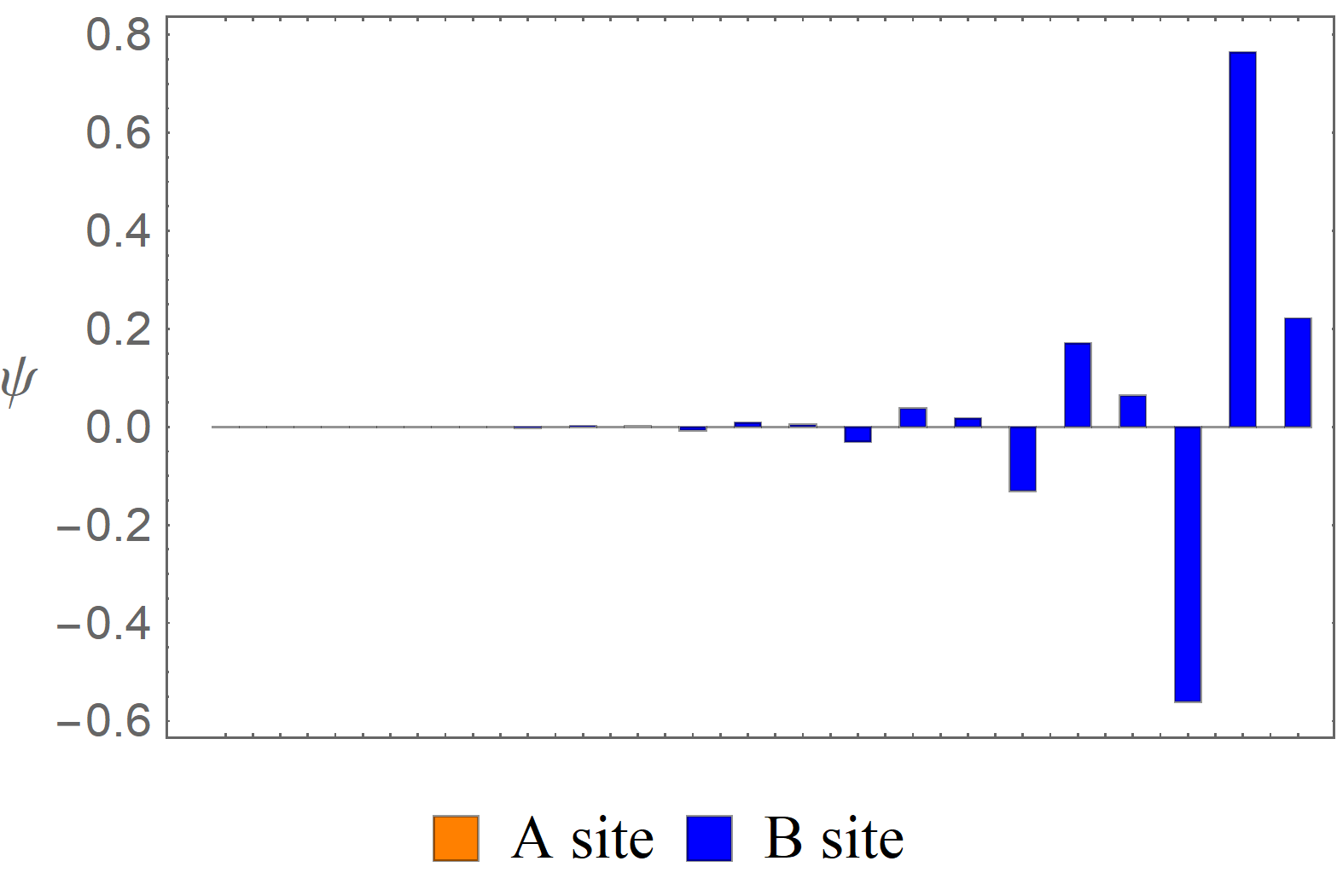}}\hskip 0.5cm
\subfloat[]{\includegraphics[width=0.30\textwidth]{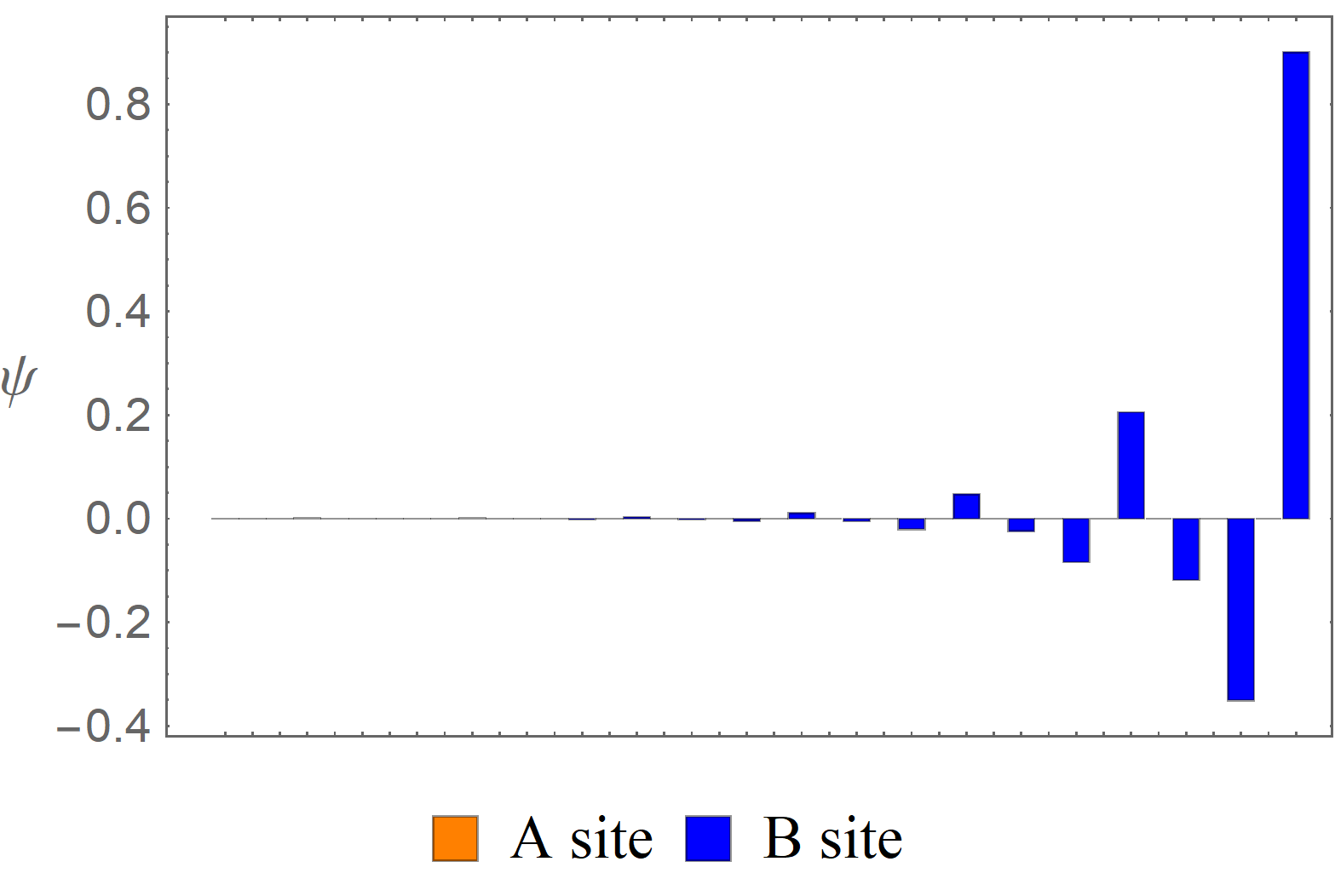}}
\caption{(a)The energy spectrum in the topological phase of the extended Rice-Mele model with $(t_0, t_1, t_2, m_0)=(3,5,8,1)$ and even number of sites so that $\n_{\rm left}=\n_{\rm right}=2$. The energies of the left and right edge states are $m_0$ and $-m_0$, respectively. (b)-(e) The wave functions of the four edge states.}\label{fig ext-RM even}
\end{figure}

\end{document}